\documentclass[12pt]{article} 
\usepackage[utf8]{inputenc}
\usepackage{geometry}
\usepackage{graphicx} 
\usepackage{dcolumn} 
\usepackage{bm} 
\usepackage[version=3]{mhchem} 
\usepackage{mathtools} 
\usepackage{soul} 
\usepackage{color} 
\usepackage[T1]{fontenc} 

\usepackage[numbers, sort&compress]{natbib}
\usepackage[hidelinks]{hyperref}
\hypersetup{colorlinks=true, allcolors=blue}

\usepackage{authblk} 

\newcommand{\affweizmann}{Department of Chemical and Biological Physics, Weizmann Institute of Science, Rehovot 7610001, Israel}
\newcommand{\affaugsburg}{Institute of Physics and Center for Advanced Analytics and Predictive Sciences, University of Augsburg, Universit\"atsstr. 1, 86159 Augsburg, Germany}
\newcommand{\affcopenhagen}{Department of Chemistry, University of Copenhagen, DK-2100 Copenhagen Ø, Denmark}
\newcommand{\affangers}{Univ Angers, CNRS, MOLTECH-Anjou, SFR MATRIX, F-49000 Angers, France}
\newcommand{\affalicante}{Departamento de Física Aplicada and Instituto Universitario de Materiales de Alicante (IUMA), Universidad de Alicante, Campus de San Vicente del Raspeig, E-03690 Alicante, Spain (current affiliation)}

\title{\textbf{Stretching helical molecular springs: the peculiar evolution of electron transport in helicene junctions}}

\author[1]{Anil Kumar Singh}
\author[2]{Yuta Ito}
\author[2]{Le\'on Martin}
\author[2]{Lukas Krieger}
\author[3]{Matea Sr\v{s}en}
\author[3]{Stephan Korsager Pedersen}
\author[4]{Axel Houssin}
\author[1]{Satyaki Kundu}
\author[1, 5]{Carlos Sabater}
\author[4]{Narcis Avarvari\thanks{narcis.avarvari@univ-angers.fr}}
\author[3]{Michael Pittelkow\thanks{pittel@chem.ku.dk}}
\author[2]{Fabian Pauly\thanks{fabian.pauly@uni-a.de}}
\author[1]{Oren Tal\thanks{oren.tal@weizmann.ac.il}}

\affil[1]{\affweizmann}
\affil[2]{\affaugsburg}
\affil[3]{\affcopenhagen}
\affil[4]{\affangers}
\affil[5]{\affalicante}

\date{} 

\begin{document}
\maketitle

\textbf{Abstract}
Single-molecule junctions represent electromechanical systems at the edge of device miniaturization. Despite extensive studies on the interplay between mechanical manipulation and electron transport in molecular junctions, a thorough understanding of conducting molecular springs remains elusive. Here, we investigate the impact of mechanical elongation and compression on the electron transport and electronic structure of helicene-based spring-like single-molecule junctions, utilizing 2,2’-dithiol-[6]helicene and thioacetyl-[13]helicene molecules bridging two gold electrodes. We observe robust, reversible U-shaped conductance variations with interelectrode distance. Ab-initio electronic structure and quantum transport calculations reveal that this behavior stems from destructive quantum interference, induced mainly by modifications of the coupling at the metal-molecule interface as a peculiar outcome of the helical backbone deformation. These findings highlight the central role of the helical geometry in combination with contact properties in the electromechanical response of conducting molecular springs, offering insights for designing functional electromechanical devices that leverage similar mechanisms.

\section{Introduction}
Single-molecule junctions are ideal systems for studying the interplay between mechanics and electronics at the atomic scale \cite{Andrea-2022, Zhang-2022}.
By varying the interelectrode distance, the shape and energy of molecular wave functions change, offering control over the electronic transport characteristics of molecular junctions. The conductance sensitivity of contacted molecules to stretching \cite{Stefani-2018, Schosser-2022, Ferri-2019, Walkey-2019, Hsu-2022, Frisenda1-2016,vanderPoel:NatComm2024}, folding \cite{Li-2021, Wu-2020} or compression \cite{Stefani-2018, Ferri-2019, Meisner-2011} has been reported, and electromechanical manipulations have been utilized for precise studies of phenomena such as quantum interference \cite{Stefani-2018, Camarasa-2020, Schosser-2022, Frisenda-2016, Reznikova-2021,vanderPoel:NatComm2024}, charge reorganization \cite{Perrin-2013} and thermoelectricity \cite{Rincon-Garcia-2016,vanderPoel:NatComm2024}.
. Since the electronic response of single-molecule junctions to mechanical forces is highly dependent on their geometry, specially designed molecules are expected to show novel and interesting electronic transport characteristics \cite{Li-2021,Schosser-2022,Hsu-2022}. In view of the scientific prospects of this research, it is important to improve our understanding of the various types of electromechanical responses of single-molecule junctions for possible use in molecule-based devices.

\begin{figure}[t!]
\centering
\includegraphics[width=1\linewidth]{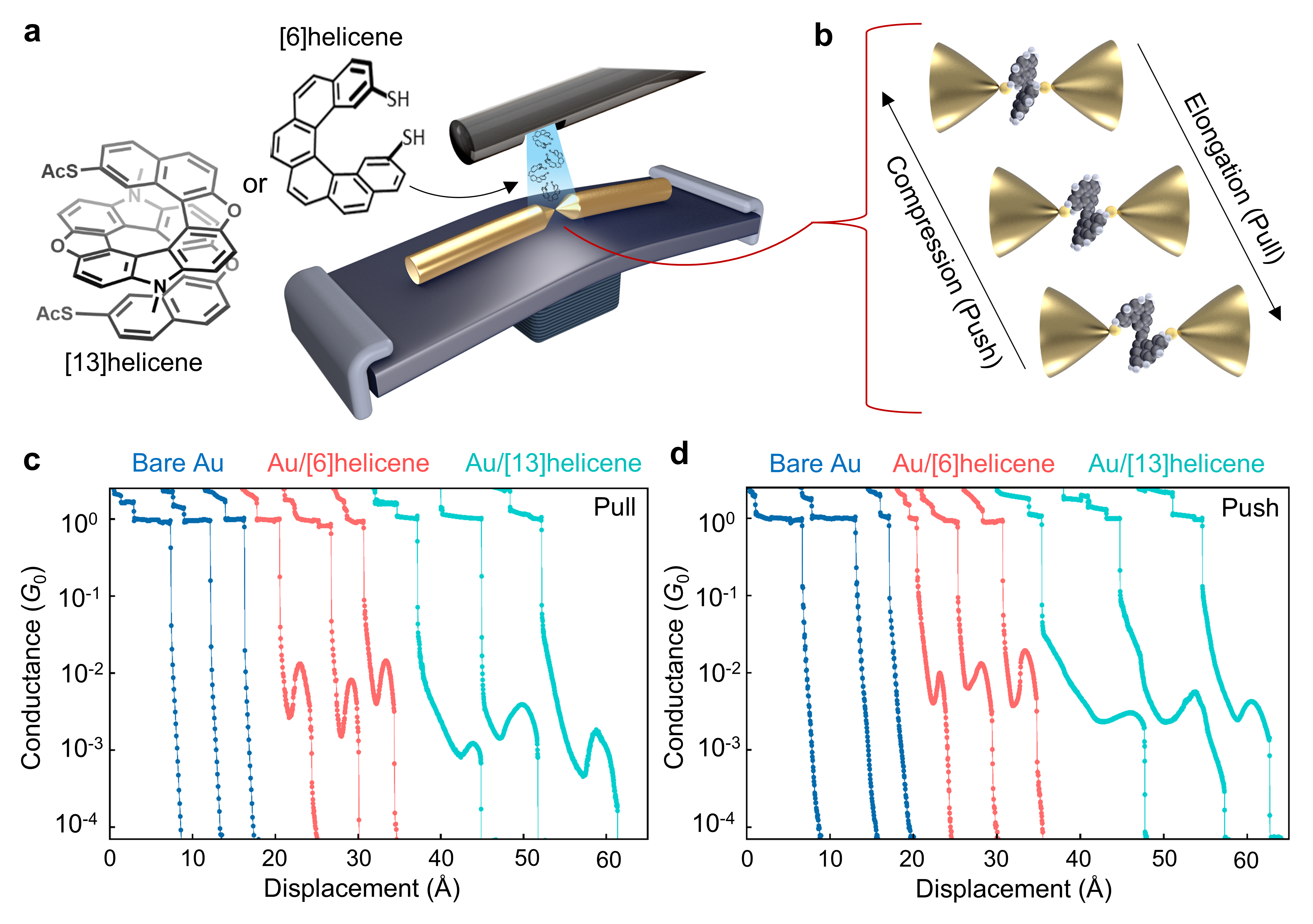}
\caption{Schematics of the break junction setup and conductance measurements. (\textbf{a}) Schematic illustration of the break-junction setup and the chemical structure of the studied helicene molecules. (\textbf{b}) Schematic illustration of pull-push cycle measurements of a single helicene molecule in the break junction. (\textbf{c}) Examples for traces of conductance versus interelectrode displacement during elongation for bare Au, Au/[6]helicene, and Au/[13]helicene junctions in light blue, red, and cyan color, respectively. (\textbf{d}) Examples of conductance traces during the compression for bare Au, Au/[6]helicene, and Au/[13]helicene junctions in light blue, red, and cyan color, respectively. All shown conductance-displacement traces are taken during elongation and compression of junctions at 200~mV applied bias voltage. The conductance is determined by dividing the current by voltage, assuming linear response.}
\label{fig:setup-Gd}
\end{figure}

The ability of a macroscale metal spring to conduct electricity is largely unaffected by its stretching state, since the length of the electron pathway and the extent of scattering remains essentially unchanged when the contact points are fixed. In contrast, the electronic properties of a molecular spring are determined by quantum mechanics. As a result, the electrical transport properties of molecular springs are more sensitive to mechanical manipulation, as will be explained further. Helicene molecules (e.g., Figure~\ref{fig:setup-Gd}a) feature a helical spring-like shape constructed from fused $\pi$-conjugated carbon rings, which may optionally be decorated with substituents \cite{Shen-2012, Pop-2019, Baciu-2020, Baciu-2022}. These molecules can therefore serve as atomic-scale conducting springs when connected to metallic electrodes in single-molecule junctions (Figures \ref{fig:setup-Gd}a,b) \cite{Treboux-1999, Guo-2015, Vacek-2015, Stetsovych-2018, Sestak-2015}. Previous theoretical studies have shown that conduction in helicene molecular junctions occurs in the off-resonant tunneling regime\cite{Ara2024}, being significantly influenced by energy shifts of molecular orbitals and changes in orbital overlap between the coils of the molecular spiral\cite{Guo-2015,Vacek-2015}. Additionally, the geometry of the metal-helicene interface has been highlighted as a key factor in determining the helicene junction's conduction properties\cite{Ara2024}. Specifically, it has been shown by calculations that through strain, intra-molecular transport may be tuned by reducing tunneling contributions between the loops of the helicene spiral for increased molecular extension \cite{Vacek-2015,Guo-2015}, concentrating transport from through space to through bond.

Here, we reveal experimentally and theoretically the effect of elongation and compression of a spring-like molecule on electrical conductance. Specifically, we study this response in different helicene molecules connected between two gold (Au) electrodes, as illustrated in Figure~\ref{fig:setup-Gd}a. In particular, we form molecular junctions based on two different helicenes: (i) a 2,2’-dithiol-[6]helicene (abbreviated as [6]helicene in the following) that has one full helical turn between the two thiol anchoring points, and (ii) a heterocyclic S-acetyl-modified [13]helicene (abbreviated [13]helicene in the following) that has two full helical turns between the two thiol anchoring points. In both cases, we find a high percentage a U-shape conductance modulation in response to changes in interelectrode distance, as shown in Figure~\ref{fig:setup-Gd}c and \ref{fig:setup-Gd}d. By U-shape we mean an initial fast conductance decrease with increasing electrode displacement, followed by an unusual increase of the conductance to a local maximum, before a final decay sets in, ending with the rupture of the molecular contact. This conductance modulation is observed both during elongation and compression of the helicene junctions. Density function theory (DFT) and quantum transport calculations explain the experimentally observed U-shape
in conductance-distance traces through characteristic changes in molecular orbital energies and molecular wave function rotations at the sulfur atoms connected to the gold electrodes, as the junction is pulled apart. In this context, we report to the best of our knowledge for the first time a peculiar switching from zero to two destructive quantum interferences inside the gap between the highest occupied molecular orbital (HOMO) and the lowest unoccupied molecular orbital (LUMO) with increasing electrode displacement. A four-level model reveals the effect as being of multiple-orbital origin. Previous simulations attributed the U-shape conductance simply to the nonmonotonic variation of the HOMO-LUMO gap with respect to displacement\cite{Guo-2015}, while our analysis uncovers a more intricate physical picture.

Overall, our work shows that spring-like helical molecular junctions with different molecules of different length and composition exhibit a similar and, in general, nonmonotonic U-shaped electromechanical response. Based on our theoretical simulations, it is the combination of the helical molecular structure and the atomic details of the binding sites that ultimately determines the appearance of this U-shape conductance behavior.

\section{Experimental results}
We use a mechanically controllable break junction setup\cite{Muller-1992} (Figure~\ref{fig:setup-Gd}a) to prepare the investigated molecular junctions in situ under cryogenic temperature (4.2 K) and cryogenic vacuum conditions. A gold wire with a notch (weak spot) at its center is attached to a flexible insulating substrate. This structure is placed in a vacuum chamber that is pumped and cooled to the setup’s base temperature by liquid helium. Bending the substrate using a piezoelectric element in a three-point bending configuration, leads to junction rupture into two freshly formed Au electrode tips. The separation between the two electrode tips can be adjusted using the piezoelement at sub-angstrom resolution, enabling the formation of a metallic atomic junction between the two electrode tips. The target molecules, either [6]helicene\cite{Anil-2024} or [13]helicene (see Section 1 of the Supporting Information for details on the synthesis of [13]helicene and Ref.~\citenum{Pedersen-2019}), are introduced from a heated local molecular source into the cold junction. This occurs during cycles of junction elongation until rupture, followed by the compression of the two tips against each other (pull-push cycles). The elongation process promotes the formation of single-molecule bridges between the electrode tips that are then stretched until rupture. The subsequent closing of the gap between the electrodes leads to a jump to molecular contact\cite{Yelin-2021}, while further junction compression eventually yields a metallic atomic contact. We use the described pull-push cycles, illustrated in Figure~\ref{fig:setup-Gd}b, to study the conductance of the molecular junctions during elongation and compression of the helicene molecular springs. Further compression between each cycle to reach a multi-atom thick contact at the narrowest cross-section, alters the atomic configuration before the next junction is formed. This step enables the analysis of helicene junctions exhibiting a variety of configurations, see section 2 of the Supporting Information.

Before introducing the helicene molecules, we analyzed the typical behavior of the bare Au atomic junctions by recording conductance-distance traces as a function of interelectrode displacement. Figure~\ref{fig:setup-Gd}c shows examples for conductance-distance traces of bare Au atomic junctions. The conductance is reduced in abrupt jumps as the number of atoms in the junction’s constriction decreases\cite{Dreher:PRB2005}. The conductance plateau at $1G_{0}$, ($G_0=2e^2/h$ being the conductance quantum, where $e$ in the elementary charge and $h$ is Planck’s constant), corresponds to a single Au atom at the smallest cross-section of the junction.\cite{Yanson-1998, Ohnishi-1998} Further elongation results in junction rupture and a conductance drop to the tunneling regime. Following the introduction of the helicene molecules, the conductance traces reveal new features, indicating the formation of molecular junctions with conductance lower than that of bare Au junctions. Figure~\ref{fig:setup-Gd}c presents typical conductance traces, measured during the elongation process after the introduction of [6]helicene or [13]helicene into the Au junctions. Interestingly, a U-shaped dip-peak feature can be observed below the typical $1G_{0}$ plateau of the Au atomic contact for both molecular junctions. Following the rupture of the junction, the electrodes are pushed against each other. Figure~\ref{fig:setup-Gd}d presents examples of conductance-distance traces, measured during the compression process, where similar U-shape features clearly appear. As a reference, we also present the conductance-distance traces for bare Au atomic junctions during the compression process, where such features are not observed. Notably, the conductance-distance traces measured for Au/[6]helicene and Au/[13]helicene during the compression and elongation processes exhibit a similar U-shape feature below $1G_{0}$. Thus, this non-monotonic variation in conductance is a general feature seen for both helicene junctions despite differences in length and atomic composition.

\begin{table}[t!]
\caption {Probability of different molecular junction features and most probable conductance.}
\label{tab:Gstatistics}
\begin{tabular}{lllll}
\hline
Junction        & Molecular traces/  & U-shape/     & U-shape/     & Most probable \\
                    & all traces         & all traces   & molecular traces & conductance \\
\hline
Au/[6]helicene & Pull: 70 $\%$ & Pull: 52 $\%$ & Pull: 75 $\%$ & $\sim 6 \times 10^{-3}G_{0}$ \\
               & Push: 82 $\%$ & Push: 64 $\%$ & Push: 77 $\%$ & $\sim 1 \times 10^{-2}G_{0}$ \\
Au/[13]helicene & Pull: 62 $\%$ & Pull: 34 $\%$ & Pull: 54 $\%$ & $\sim 2 \times 10^{-3}G_{0}$ \\
                    & Push: 70 $\%$ & Push: 44 $\%$ & Push: 62 $\%$ & $\sim 1 \times 10^{-2}G_{0}$ \\
\hline
\end{tabular}
\begin{minipage}{\textwidth}
\vspace{0.2cm}
Second column: percentage of traces with distinctive molecular conductance features (not necessarily a U-shape) below $1G_{0}$, out of the total number of traces. Third column: percentage of traces, showing a U-shape feature, out of the total number of traces. Fourth column: percentage of traces, showing a U-shape feature, out of all molecular traces. Fifth column: most probable conductance of the respective molecular junction, defined as the conductance most frequently observed below $1G_{0}$ (i.e., the conductance of the predominant peak below $1G_{0}$, see {Figure S5} of the Supporting Information). The values presented are based on 20,000 pull and 20,000 push conductance-distance traces for each of the two junction types.
\end{minipage}

\end{table}

To study the general properties of the U-shape feature in an ensemble of molecular junctions, we analyzed 20,000 pull-push cycles for each molecular junction type. As presented in Table~\ref{tab:Gstatistics}, among the traces collected during both elongation (pull) and compression (push) of Au/[6]helicene and Au/[13]helicene junctions, most traces exhibit molecular features -- specifically, conductance features below $1G_{0}$ that deviate from the monotonous tunneling decay but do not necessarily exhibit a U-shape. The formation probability of molecular junctions based on [6]helicene is consistently higher than that of [13]helicenes, both in the pull and push processes. Among the successfully formed molecular junctions, the majority shows a U-shape feature.

\begin{figure}[t!]
\centering
\includegraphics[width=1\linewidth]{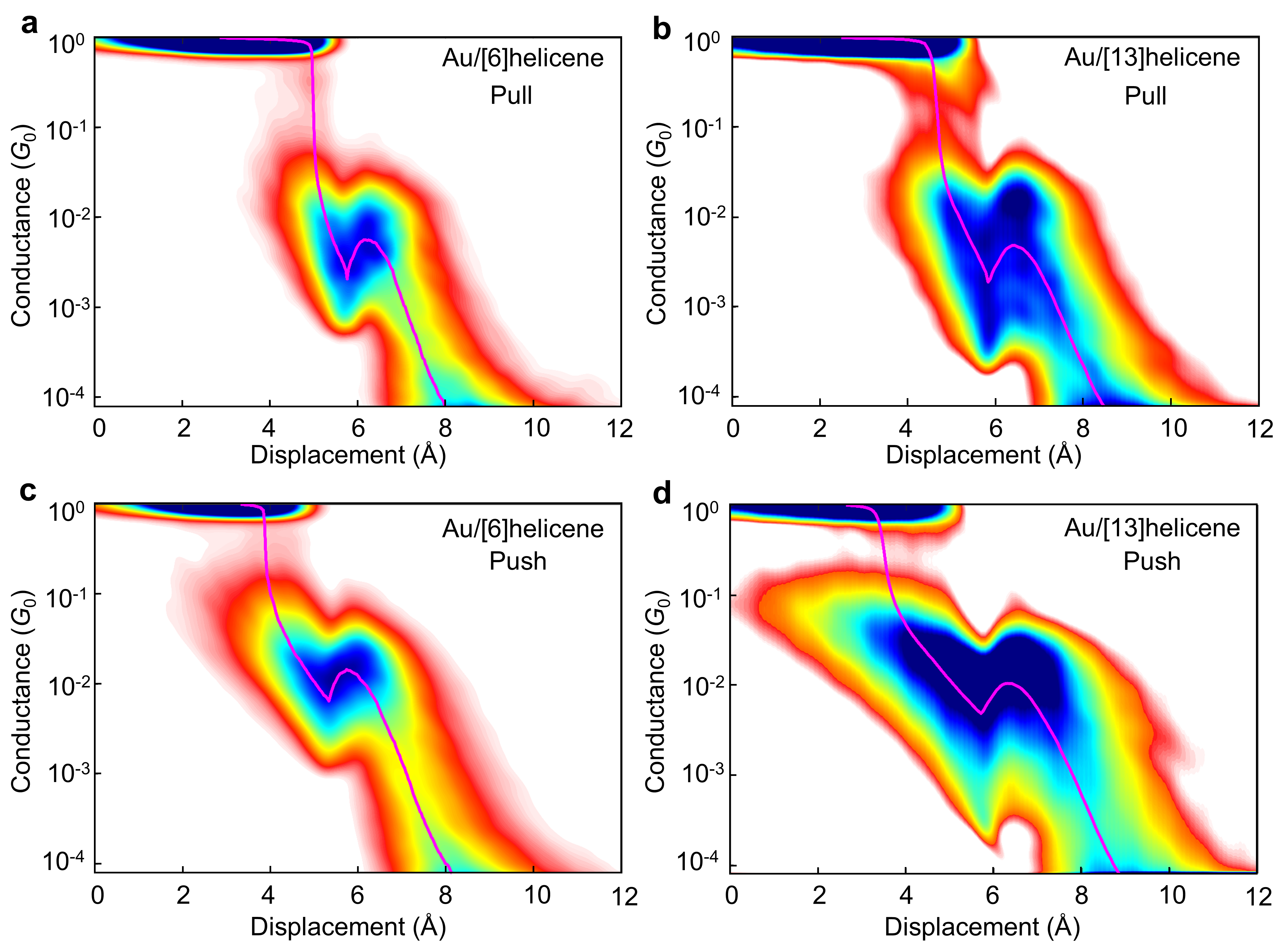}
\caption{Conductance-displacement density plot for molecular junctions during pull and push processes. Conductance-displacement density plot of (\textbf{a}) Au/[6]helicene and (\textbf{b}) Au/[13]helicene molecular junctions during pulling. (\textbf{c},\textbf{d}) Same as panels (a) and (b) but for pushing.
}
\label{fig:2Dhistos}
\end{figure}

In Figure~\ref{fig:2Dhistos}, we present density plots of U-shaped conductance traces as a function of interelectrode displacement for both molecular junction types. The traces are aligned to the conductance minimum, revealing a high-density region with a “heart” shape. The overlayed median trace also clearly reflects the U-shape. In contrast to the individual traces, displayed in Figure~\ref{fig:setup-Gd}c and \ref{fig:setup-Gd}d, here the average minimum is sharper. This results from the alignment to the minima, and if the traces are instead aligned to the maxima of the U-shape, the average maximum is sharper whereas the average minimum is more rounded, see Section 2.4 of the Supporting Information.

\begin{figure}[t!]
\centering
\includegraphics[width=1\linewidth]{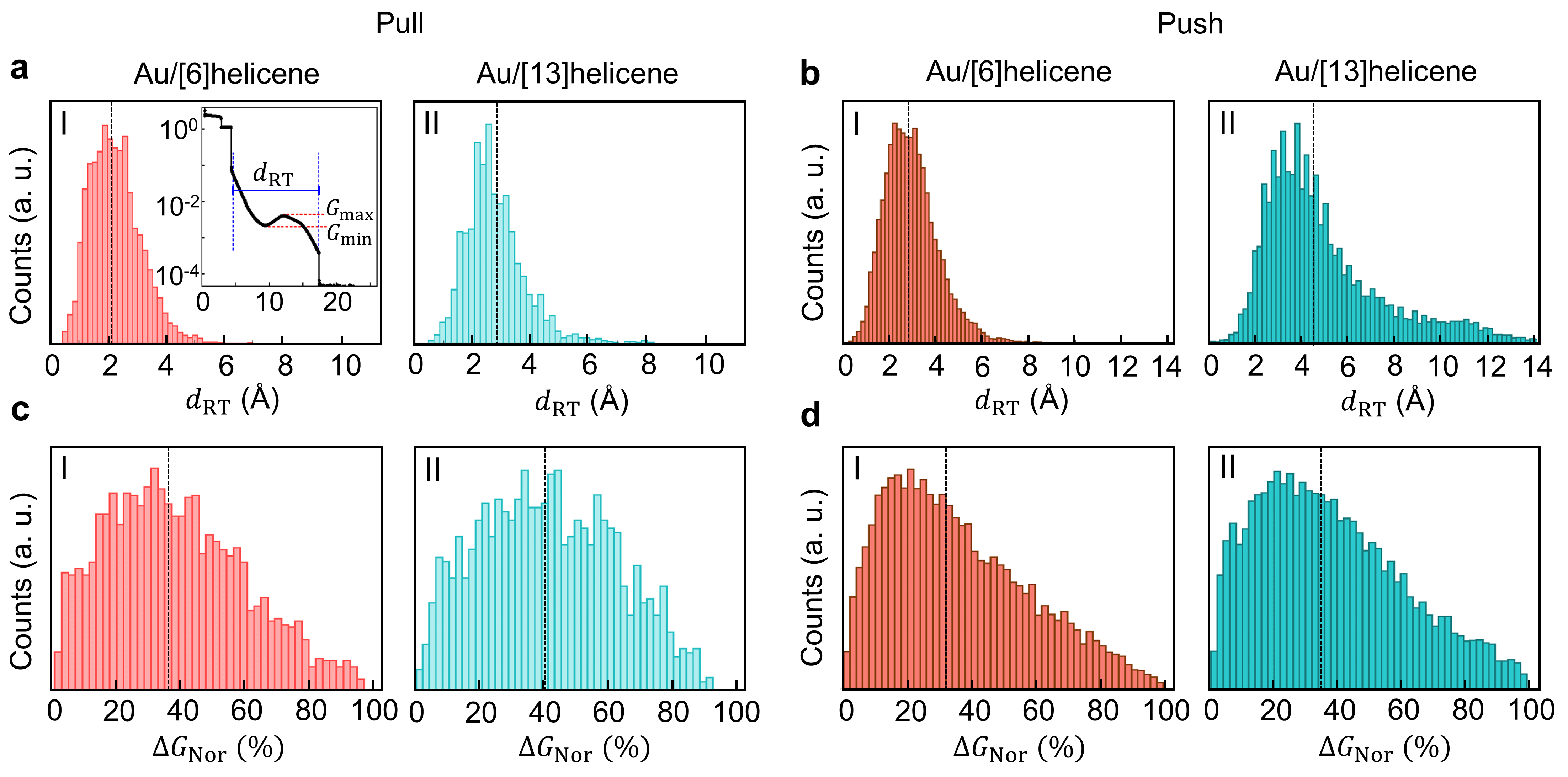}
\caption{One-dimensional histogram of the U-shape displacement $d_\text{U}$ and percental relative upturn conductance change. (\textbf{a}) One-dimensional histogram of $d_\text{U}$, constructed from pull traces of Au/[6]helicene (I) and Au/[13]helicene (II) junctions. The inset in panel (\textbf{a},I) illustrates the definitions of $d_\text{U}$, $G_\text{max}$ and $G_\text{min}$ for an example conductance trace exhibiting a U-shape. (\textbf{b}) The same as panel (a) but for push traces. (\textbf{c}) One-dimensional histogram of $\Delta G_\text{Nor}$, constructed from pull traces of Au/[6]helicene (I) and Au/[13]helicene (II) junctions. (\textbf{d}) The same as panel (b) but for push traces. The medians of $d_\text{U}$ and $\Delta G_\text{Nor}$ histograms are marked by dashed lines.}
\label{fig:dU-dG-ana}
\end{figure}

Additionally, we analyzed the U-shape feature of the measured conductance traces, as shown in Figure~\ref{fig:dU-dG-ana}. We determined its length ($d_\text{U}$), starting from the Au junction rupture point up to the molecular junction rupture, and we further quantified the relative upturn molecular conductance change ($\Delta G_\text{Nor} = [(G_\text{max}-G_\text{min})/(G_\text{max}+G_\text{min})]$), each based on individual traces. The inset of Figure~\ref{fig:dU-dG-ana}a(I) depicts $d_\text{U}$, $G_\text{max}$ and $G_\text{min}$ for an example conductance trace with a U-shape feature. Figure~\ref{fig:dU-dG-ana}a(I-II) and \ref{fig:dU-dG-ana}b(I-II) present $d_\text{U}$ histograms, generated from pull and push traces for the two molecular junction types, together with the respective median values. The median displacement for Au/[13]helicene junctions is approximately 1.3 times higher than that for Au/[6]helicene junctions during the elongation process and approximately 1.6 times higher during the compression process. These ratio is in good agreement with the relative length of the molecules, as found by DFT calculations (see below for details). Specifically, the gas-phase distance between the outer sulfur atoms of each molecule type is 5.5~\AA\ and 7.7~\AA\ for the [6]helicene and [13]helicene molecules, respectively, resulting in a length ratio of 1.38 in the relaxed configuration. For a fully stretched junction near rupture, the calculated sulfur-sulfur distances for the stretched [6]helicene and [13]helicene molecules are 10.73~\AA\ and 14.71~\AA, respectively, for one configuration (TT, see below), and 11.54~\AA\ and 15.79~\AA\ for another configuration (HH, see below). This results in a length ratio of 1.37 for both configurations in fully extended junctions.

$\Delta G_\text{Nor}$ histograms, extracted from the pull and push traces for the two helicene junctions, are shown in Figure~\ref{fig:dU-dG-ana}c and \ref{fig:dU-dG-ana}d. Notably, histograms for both types of molecular junctions exhibit qualitatively similar shapes in the pull and push processes. This similarity suggests that the upturn conductance behavior is a general characteristic of the examined helicene molecules. The histograms display a distinct single peak, indicating that the change in conductance, $G_\text{max}-G_\text{min}$, generally correlates with the average conductance value, ($G_\text{max}+G_\text{min}$)/2, at which the U-shape appears. We therefore propose that variations in the U-shape height primarily result from variations in the overall conductance of the junction across different junction realizations, while the effect of elongation or compression itself has a rather robust influence on the conductance, within the subset of molecular junctions that show a U-shaped conductance.

\section{Theoretical results}
To understand the mechanism behind the U-shape conductance variations we turn to ab-initio calculations. Electron transport through helicene single-molecule junctions has been studied theoretically in previous reports\cite{Vacek-2015,Guo-2015,Ara2024}. Although a U-shape variation in conductance as a function of interelectrode distance has been predicted\cite{Guo-2015}, the present understanding is still rather limited with regard to conditions, under which such a behavior occurs and how it originates from molecular properties. To improve this situation and explain the experimental observations, we conducted quantum transport simulations based on DFT.

DFT calculations, employing the quantum chemistry program suite TURBOMOLE\cite{TURBOMOLE2023}, were used to determine the electronic structure and minimum energy geometries of helicene single-molecule junctions. Based on this, electronic transmission and conductance were calculated from Landauer-B\"uttiker scattering theory, expressed in terms of nonequilibrium Green’s function techniques using our own code\cite{pauly2008cluster}. Since experiments are conducted at low temperatures, we determine the electrical conductance for different electrode displacements $d$ from the electronic transmission via $G(d)=G_0\tau(E_\text{F},d)$ with the transmission $\tau(E,d)$, the Fermi energy $E_\text{F}$ and the conductance quantum $G_0$. Details of computational settings are described in section 3.1 to 3.3 of the Supporting Information.

\begin{figure}[t!]
\centering
\includegraphics[width=1\linewidth]{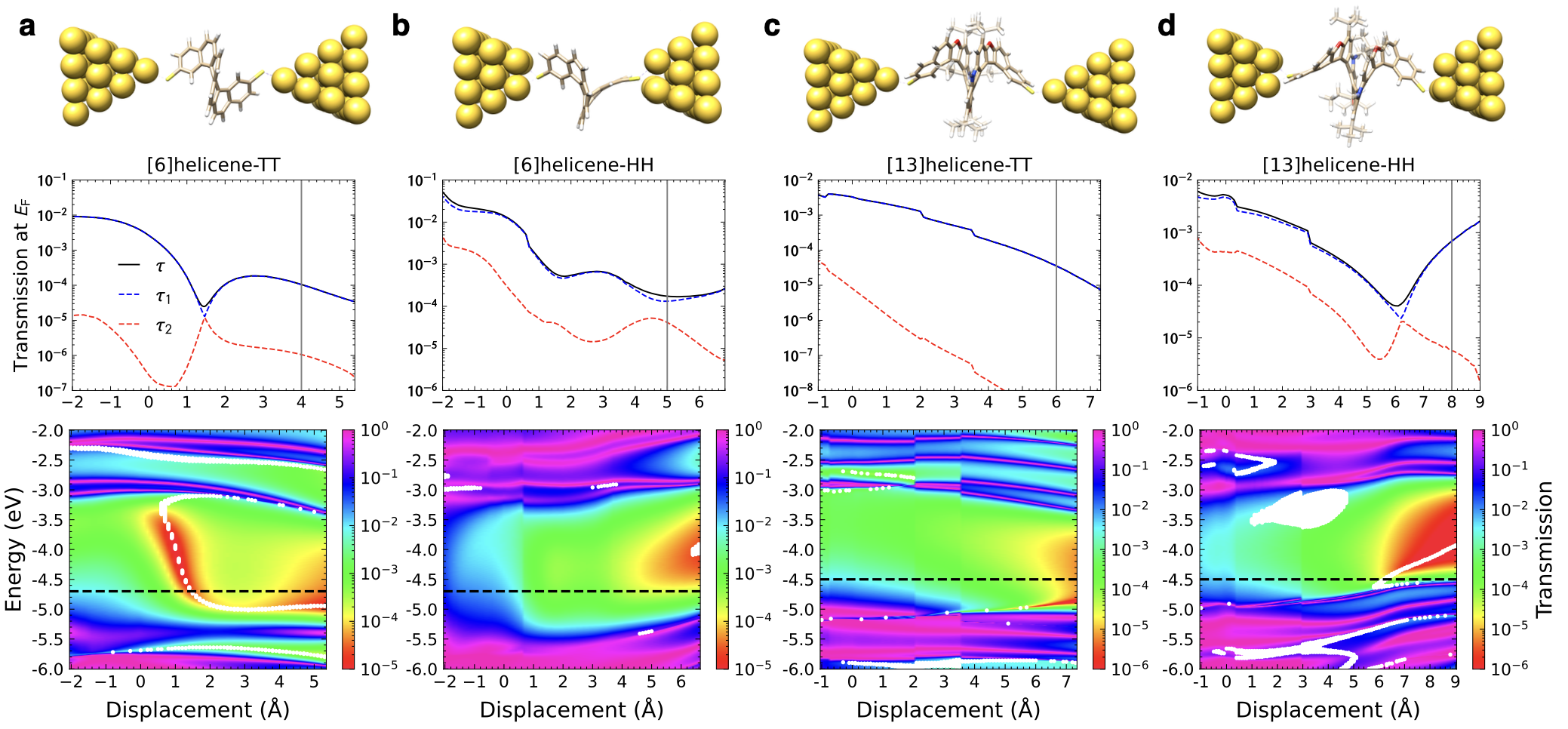}
\caption{Results of DFT-based quantum transport simulations for different helicene single-molecule junctions, namely (a) [6]helicene-TT, (b) [6]helicene-HH, (c) [13]helicene-TT and (d) [13]helicene-HH. The first row shows the simulated helicene junction at a selected electrode separation. The second row shows the total transmission $\tau(E_\text{F},d)$ and the transmission of the first and second eigenchannel, $\tau_{1}(E_\text{F},d)$ and $\tau_{2}(E_\text{F},d)$, as a function of the electrode displacement $d$ at $E_\text{F}$. The gray vertical line indicates the displacement at which the junction geometries are plotted in the first row. The third row shows the transmission as a function of energy and displacement, $\tau(E,d)$. Points, where the transmissions of first and second eigenchannels are similar ($\tau_{2}(E,d)/\tau_{1}(E,d)\ge 0.7$), are indicated by white dots in each two-dimensional transmission map. Horizontal dashed black lines indicate the Fermi energy $E_\text{F}$. Note that all transport properties are shown up to the rupture point, i.e.\ contacts break in the next pulling step of $0.1$~\AA\ at the largest displacements.}
\label{fig:DFT_transmission}
\end{figure}

To explore the geometric variability of the distance-dependent conductance, we study two types of single-molecule junctions for [6]helicene and [13]helicene, respectively (see examples in Figure ~\ref{fig:DFT_transmission}, top panel). A starting geometry is built by placing the corresponding helicene molecule between two gold electrodes. The gold electrodes are of pyramidal-shape, featuring either two atomically sharp tips in the top-top (TT) junctions or blunt tips in the hollow-hollow (HH) junctions, where the tip atoms on both sides are removed. These junctions are then stretched or compressed, and energies of atomic positions are optimized in each electrode displacement step under the constraint of a given junction length, measured as the distance between fixed outer gold atomic layers. Further details on the molecular junction construction and the elongation and compression procedures can be found in section 3.2 of the Supporting Information.

To compare to the experimental conductance, the total transmission at the Fermi energy and its decomposition into the most transmissive
eigenchannels is shown in figure~\ref{fig:DFT_transmission} as a function of the electrode displacement. Due to the proportionality between $G(d)$ and $\tau(E_\text{F},d)$ we use both synonymously in the following. The [13]helicene-TT junction exhibits a rather linear decay of conductance on the logarithmic scale, interrupted by some discontinuous changes due to geometric rearrangements. The more typical behaviour in our simulations, however, is a nonmonotonous conductance dependence on distance, visible for [6]helicene-TT, [6]helicene-HH and [13]helicene-HH junctions. For the [6]helicene-HH junction, two local conductance minima are found, but they are rather shallow. For [6]helicene-TT and [13]helicene-HH junctions the single dips are very pronounced, resembling the experimental conductance-distance trace in figure~\ref{fig:setup-Gd}. In both cases the transmissions of the eigenchannels 1 and 2 are degenerate near the respective local conductance minimum.
In section 3.4 of the Supporting Information, we analyze the wave functions of the first and second transmission eigenchannel and show that the channels exchange their character, meaning that the conductance minima can be seen as "eigenchannel crossing points".
The results of figure~\ref{fig:DFT_transmission} demonstrate that the distance dependence of the conductance of the helicene molecular junctions is sensitive to the geometry of the gold electrodes. These findings are in agreement with the experimental observation that a U-shape conductance occurs only in a subset of the ensemble of examined molecular junctions.

In order to get a better overview of the transport properties, we show the transmission $\tau(E,d)$ as a two-dimensional contour plot in dependence of energy and electrode displacement at the bottom panel of figure~\ref{fig:DFT_transmission}.
The purple-colored areas of high transmission show how transmission resonances evolve and thus indicate the displacement dependence of orbital energies.
A suppressed transmission, as signaled by the red-shaded areas, is typically found at the end of the pulling process. Of particular interest is the transmission valley visible for the [6]helicene-TT junction at small electrode displacements, causing a fast drop of the transmission at the Fermi energy
and subsequent increase at higher $d$ (see second and third rows of figure~\ref{fig:DFT_transmission}a).
To explore the behavior of eigenchannel transmissions, we mark points $(E,d)$, where the transmissions of first and second channels are similar within the accuracy $\tau_{2}(E,d)/\tau_{1}(E,d)\ge 0.7$, in white in the maps of figure~\ref{fig:DFT_transmission}. This reveals interesting arc structures for the junctions [6]helicene-TT and [13]helicene-HH, which exhibit the pronounced local conductance minima.

\begin{figure}[t!]
\centering
\includegraphics[width=0.9\linewidth]{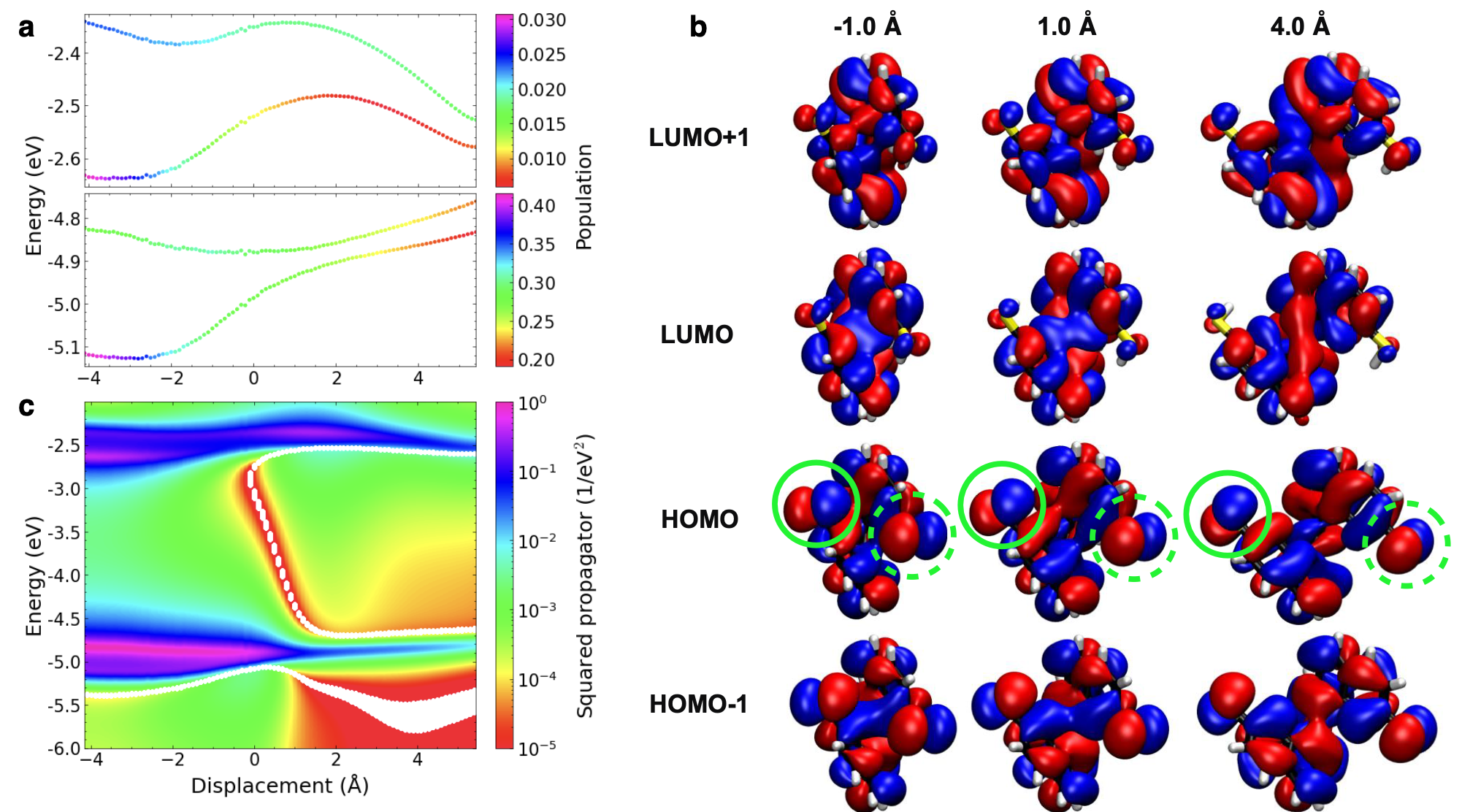}
\caption{(a) Evolution of the energies of frontier levels HOMO-1, HOMO, LUMO, LUMO+1 as a function of electrode displacement. They are determined from DFT calculations of the [6]helicene-TT junction by removing all gold atoms and saturating terminal sulfur atoms at each side with a single hydrogen. Color-coded for each orbital is the change in Mulliken population on a sulfur atom. (b) wave functions of the four frontier molecular orbitals at displacement points $d=-1.0$, 1.0 and 4.0~\AA, as obtained from DFT. For the HOMO, terminal sulfur atoms are encircled in green to show that the $\pi$-symmetric orbitals on these atoms change their orientation upon stretching. (c) Two-dimensional contour plot of the square of the absolute value of the retarded propagator, $|G^{(0),\text{r}}_\text{lr}|^2$, as a function of energy and electrode displacement for the four-level model. The energy levels are taken from DFT calculations, see panel (a), and other parameters such as residues and the broadening $\eta=0.06$~eV are chosen to mimic the DFT results of the [6]helicene-TT junction in figure~\ref{fig:DFT_transmission}a. We normalize the squared propagator such that the maximum value inside the full energy-distance map equals 1. See section 3.5 of the Supporting Information for further discussion and choice of parameters.}
\label{fig:4level-model}
\end{figure}

Let us now analyze at the example of the [6]helicene-TT junction, which ingredients lead to the pronounced dip in $\tau(E_\text{F},d)$. The evolution of HOMO-1, HOMO, LUMO and LUMO+1 energies as a function of electrode displacement is plotted in figure~\ref{fig:4level-model}a. The energetic changes of molecular orbital energies have been argued to cause the dip structure \cite{Guo-2015}, but for this particular junction, the changes in energy are moderate. In particular no crossing of levels is found, but level energies just approach each other at large displacements.

To study modifications in the molecular orbitals, we extracted the molecular part from the [6]helicene-TT junctions and added single hydrogen atoms at the sulfur anchors. The color-coded Mulliken charges on the terminal sulfur atoms change substantially with the extension of the helicene, as visible in figure~\ref{fig:4level-model}a. Note that (HOMO-1,HOMO) and (LUMO,LUMO+1) are plotted separately due to their difference in energy and population range. The Mulliken analysis indicates crucial modifications in the orbital character at the anchors as a function of the molecular extension.

The four frontier orbitals of the molecular part of the [6]helicene-TT junction are displayed in figure~\ref{fig:4level-model}b at the three different displacement points, $d=-1.0$, 1.0, and 4.0~\AA. We find that the $\pi$-like orbitals on the sulfur atoms, indicated by the green circles, change their orientation, as the helicene is stretched. With respect to a transport from left to right, this changes the orbital character from initially $\sigma$-like at small displacement to finally $\pi$-like at large displacement, but the precise orbital character of the connection to the gold electrodes will of course depend on the detailed junction geometry. The small Mulliken population of the LUMO and LUMO+1 in figure~\ref{fig:DFT_transmission}a is consistent with the low weight of LUMO and LUMO+1 wave functions for these levels in figure~\ref{fig:DFT_transmission}b.

The appearance of a destructive interference valley in figure~\ref{fig:DFT_transmission}a for the [6]helicene-TT junction is surprising. Considering only the HOMO and LUMO states in figure~\ref{fig:4level-model}b, no destructive quantum interference should occur due to unchanged but opposite parities of these orbitals for all displacements according to the orbital rule for transport through molecules\cite{Yoshizawa-2012}.
The peculiar arc structure of transmission eigenchannel degeneracies in figure~\ref{fig:DFT_transmission}a reveals that we observe for this junction a transition from zero to two destructive quantum interferences in the HOMO-LUMO gap, as the helicene molecule is stretched. The destructive quantum interference at higher energies is hidden, however, by the LUMO-related transmission resonances.

To explain the transmission map of figure~\ref{fig:DFT_transmission}a, we study a four-level model, taking the HOMO-1, HOMO, LUMO and LUMO+1 into account. The details of our toy model are explained in section 3.5 of the Supporting Information. We approximate the transmission by the square of the absolute value of the molecule-internal propagator as $\tau(E,d)/\gamma^2 \approx |G^{(0),\text{r}}_\text{lr}(E,d)|^2$, with the zeroth order retarded Green's functions of the molecule, $G^{(0),\text{r}}_\text{lr}(E)=\sum_{m=\text{HOMO-1}}^\text{LUMO} \rho_{\text{lr},m}(d)/(E-\epsilon_{m}(d)+i\eta)$. Here, $\rho_{\text{lr},m}(d)$ is the residue of orbital $m$, given by the product of orbital expansion coefficients at the sulfur atoms on the left (l) and right side (r), $\epsilon_{m}(d)$ is the orbital energy, $\gamma$ is a scalar parameter characterizing the coupling of the molecule to the gold electrode, and $\eta$ is an infinitesimal broadening parameter.

A transmission map calculated from the four-level model is shown in figure~\ref{fig:4level-model}c. We used $\epsilon_m(d)$ from the DFT calculations in figure~\ref{fig:4level-model}a and adjusted the residues $\rho_{\text{lr},m}(d)$ to generate a transmission map similar to figure~\ref{fig:DFT_transmission}a. We note that to reproduce the transmission map, the distance dependence of the molecular orbital energies $\epsilon_m(d)$ is not essential but the variation of the residues $\rho_{\text{lr},m}(d)$.
White dots in figure~\ref{fig:4level-model}c show the points, where $G^{(0),\text{r}}_\text{lr}(E,d)$ vanishes, if the infinitesimal broadening $\eta$ is set to zero. We indeed find the expected transition from zero to two destructive quantum interferences inside the HOMO-LUMO gap. The one at higher energies is however masked by the LUMO-related transmission resonances.

In summary, our theoretical analysis shows that it is not sufficient to consider only the displacement dependence of molecular orbital energies or the parity of the wave functions of HOMO and LUMO on anchor atoms to explain the electronic transport properties of flexible, helical helicene single-molecule junctions. Instead, it is crucial to take the distance dependence of molecular orbital wave functions on the terminal sulfur atoms into account, see figure~\ref{fig:4level-model}, that arises from changes in orbital orientation with increasing helicene extension.

This analysis explains why the distance dependence of the conductance for helicene single-molecule junctions is sensitive to the molecule-electrode interface. We furthermore reveal a peculiar change from zero to two destructive quantum interferences as a multiorbital effect. These findings are expected to be relevant for charge transport studies with other flexible molecules, such as extended mechanoelectrically sensitive $\pi$-stacked molecules\cite{Schosser-2022,Hsu-2022}, a research field of large present interest.

\section{Conclusions}

Our study shows that a suppression and subsequent revival of the conductance for increasing electrode displacement are a characteristic of spring-like molecules as helicene when contacted by gold metal electrodes.
The transition from zero to two destructive interferences in the HOMO-LUMO gap and the control of the energetic position of destructive quantum interferences with distance makes the helical helicene a unique class of mechanosensitive molecules. The finding of mechanical control of electron transport through helical molecules lays the basis for further studies, where thermoelectric,
spin or heat transport through molecular nanostructures might be tuned by mechanical modifications of helical molecules. Even mechanical manipulations of current-induced forces acting on a helical molecular rotor can be envisioned, offering precise control over nanoscale devices beyond what can be achieved with bulk materials. These examples highlight the potential of strained helical molecules in advancing the design and functionality of next-generation miniaturized nanoelectronic molecular systems.

\section{Methods}

\subsection{Experimental}

Experiments utilized a mechanically controllable break-junction (MCBJ) setup (Fig. 1a). A gold wire (99.998$\%$, 0.1 mm, Alfa Aesar) was attached to a phosphor-bronze substrate (1 mm thick) coated with a 100 $\mu$m insulating Kapton film, and assembled in a vacuum chamber cooled with liquid helium. A three-point bending mechanism broke the wire at a central notch, creating a gap adjustable to sub-Ångström precision using a piezoelectric element (PI P-882 PICMA) controlled by a 24-bit NI-PXI4461 data acquisition (DAQ) card.

Conductance traces of bare gold junctions were recorded as a function of electrode displacement to confirm typical Au contact conductance before introducing target molecules. The racemic mixture of [6]helicene (2,2'-dithiol-[6]helicene) was synthesized as per Ref.~\citenum{Anil-2024}, while [13]helicene (SAc[13]SAc) was prepared according to Section 1 of the Supporting Information. Molecules were sublimated into the cold junction using a heated source during repeated pull-push cycles, stopping once conductance deviated from the typical Au signature. Molecular junctions were analyzed by forming contacts at $\sim50$ to $70G_0$ and stretching until rupture at 10–20 Hz. Conductance was measured under a 200 mV bias to enhance the signal-to-noise ratio, with the current amplified (Femto DLPCA 200) and recorded by the DAQ card at a sampling rate of 100–200 kHz. Conductance values were calculated by dividing current by voltage, with interelectrode displacement estimated based on the exponential dependence of tunneling currents on electrode separation, as detailed in Refs.~\citenum{Tamar-2013, Sudipto-2022}.

\subsection{Theoretical}

Density functional theory (DFT) calculations were conducted using the TURBOMOLE\cite{TURBOMOLE2023} package to analyze the electronic structure and optimized geometries of helicene-based single-molecule junctions, employing the PBE \cite{Perdew1996} exchange-correlation functional and def-SVP\cite{Schafer1992} basis set, with dispersion corrections\cite{disp3} for van der Waals interactions. Energy convergence was set to $10^{-8}$~a.u, with geometry optimizations halting when Cartesian gradients fell below $10^{-5}$~a.u.

Single-molecule junctions were modeled via an extended central cluster (ECC) approach, integrating helicene molecules and portions of gold electrodes \cite{pauly2008cluster}. Two electrode configurations were explored: atomically sharp top-top (TT) and blunt hollow-hollow (HH) junctions. ECCs were stretched or compressed in 0.1~\AA\ increments while optimizing geometries until rupture or non-convergence occurred.

Electronic transmission and conductance were calculated using Landauer-Büttiker theory within the nonequilibrium Green’s function (NEGF) formalism \cite{pauly2008cluster}. The conductance, $G=G_0\tau(E_\text{F})$, was assessed at low temperatures ($T\approx 4.2$~K), where $G_0=2e^2/h$ is the conductance quantum and $\tau(E_{\text{F}}, d)$ represents transmission at the Fermi energy, utilizing a mesh of $32 \times 32$ transverse $k$-points for high precision \cite{pauly2008cluster}. To address DFT energy level alignment limitations, a DFT+$\Sigma$ correction \cite{dfts_neaton,dfts_zotti} was applied to selected geometries, and we adjusted the Fermi level subsequently to align better with these corrected calculations.

\section{Author Contribution}

The project was conceived by A.K.S. and the corresponding authors. M.S. and S.K.P. performed the chemical synthesis, purification, and characterization of the [13]helicenes under the supervision of M.P. A.H. synthesized the [6]helicenes under the supervision of N.A. A.K.S. fabricated the studied atomic and molecular junctions, and performed the experiments, under the supervision of O.T. C.S. conducted the initial conductance measurements of [13]helicene-based junctions. The experimental data analysis was performed by A.K.S. and S.K., with guidance from O.T. Y.I., L.M., and L.K. carried out the theoretical modeling, including DFT calculations and their analysis, under the supervision of F.P. All authors contributed to the preparation and writing of the manuscript.

\section{Supporting Information}
The synthesis and characterization of all newly developed compounds, experimental details, characterization of atomic and molecular junctions, examples of conductance traces as a function of displacement for molecular junctions displaying a U-shaped feature, additional analysis details,
further details on DFT and quantum transport calculations.

\section*{Acknowledgments} 

O.T.\ appreciates the support of the Harold Perlman family and acknowledges funding by research grants from Dana and Yossie Hollander, the Ministry of Science and Technology of Israel (grant number 3-16244), Israeli Science Foundation (Grant 2129/23), and the European Research Council, Horizon 2020 (grant number 864008). F.P.\ is supported through the Collaborative Research Center 1585, project C02 of the German Research Foundation (DFG), grant number 492723217. He and his group furthermore gratefully acknowledge the resources provided by the LiCCA high-performance computing cluster of the University of Augsburg, co-funded by the DFG, grant number 499211671. M.P. is grateful for financial support from the Danish Council for Independent Research (DFF 4181-00206 and 9040-00265) and from the University of Copenhagen. A.H. and N.A. gratefully acknowledge support in France by the CNRS, the University of Angers and the French National Agency for Research (ANR) project SECRETS (ANR PRC 20-CE06-0023-01).

\bibliography{bibliography}

\end{document}


\maketitle

\setcounter{figure}{0}
\renewcommand{\thefigure}{S\arabic{figure}} 

\tableofcontents

\section{Molecule synthesis and characterization}

\subsection{Synthesis} A bis-thioacetyl (AcS[13]SAc) derivative of the longest-reported diazatrioxa[13]helicene, \cite{Pedersen-2019} was prepared for this study. The bis-hydroxy[13]helicene (HO[13]OH) was transformed to the bis-thiol[13]helicene (HS[13]SH) through a Newman-Kwart rearrangement. In the final synthetic step of the synthetic protocol, the dimethylcarbamoyl protecting group was removed by treatment with n-BuLi (See the Supporting Information, SI), and the formed doubly deprotected dithiol analogue of [13]helicene contained the free thiolates. The reaction was quenched with acetyl chloride to generate the thioacetyl groups and prevent the unwanted oxidation of thiols. The final product is AcS[13]SAc.

\subsection{General Experimental Procedures}
All chemicals, unless otherwise stated, were purchased from commercial suppliers, and used as received. All solvents were high-performance liquid chromatography (HPLC) grade. Analytical thin-layer chromatography (TLC) was performed on SiO$_2$ 60 F254 0.2 mm-thick precoated TLC plates. Flash column chromatography was performed using silica gel (Silica gel 60 (43‒60 $\mu$m) purchased from VWR).

$^1$H NMR and $^{13}$C NMR spectra were recorded at 500 and 126 MHz, respectively, using residual non-deuterated solvent (CDCl$_3$ $\delta$H = 7.26 ppm and $\delta$C = 77.16 ppm; DMSO-d$_6$ $\delta$H = 2.50 ppm and $\delta$C = 39.52 ppm; CD$_2$Cl$_2$ $\delta$H = 5.32 ppm and $\delta$C = 53.84 ppm as the internal standard). All chemical shifts ($\delta$) are quoted in ppm, and all coupling constants (J) are expressed in hertz (Hz). The following abbreviations are used for convenience in reporting the multiplicity for NMR resonances: s = singlet, d = doublet, t = triplet, dd = doublet of doublets, and m = multiple.

High-resolution mass spectrometry (HRMS) spectra were recorded on an ESP-MALDI-FT-ICR instrument equipped with a 7T magnet (prior to the experiments, the instrument was calibrated using NaTFA cluster ions).

\subsection{Synthesis of thioacetyl[13]helicene}

\begin{figure}[t]
\centering
\includegraphics[width=1\linewidth]{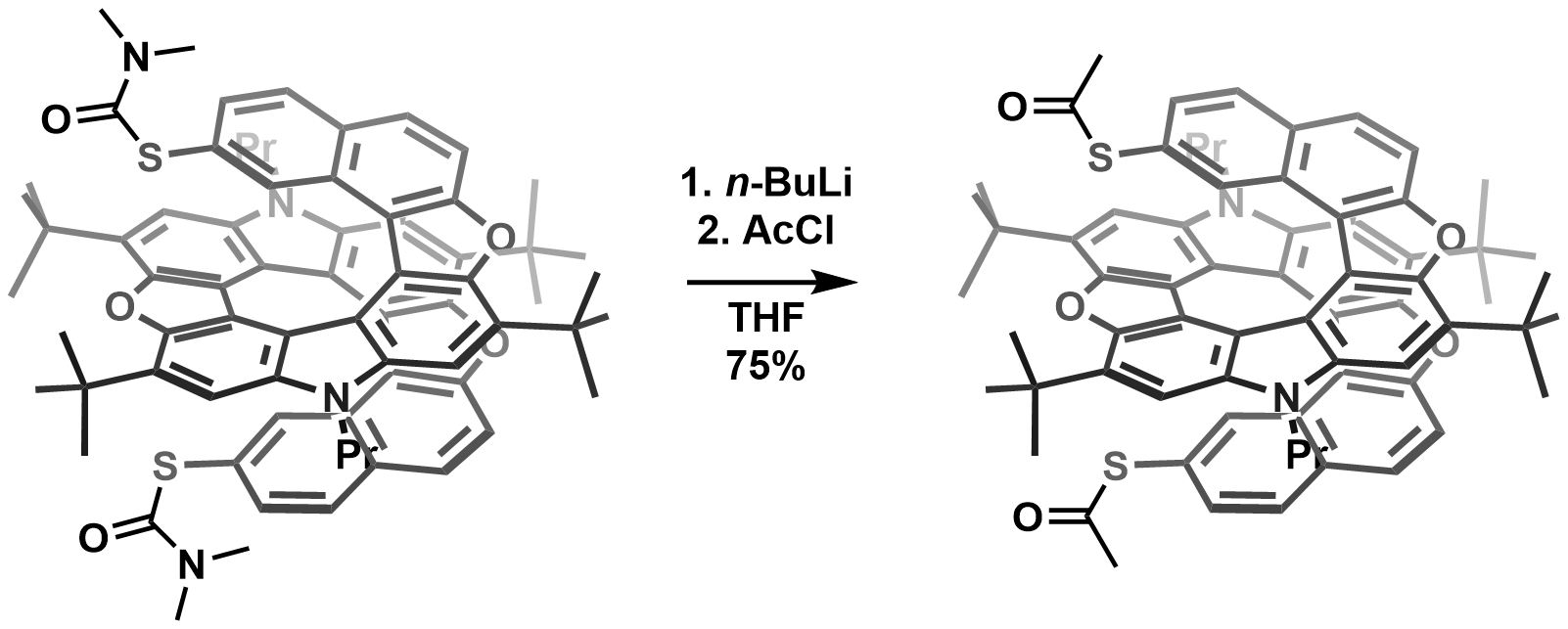}
\caption{Synthesis of SAc-functionalized [13]helicene.}
\label{fig:synthesis1}
\end{figure}

Starting material (15 mg, 0.013 mmol, 1.0 eq.) was added to a flame dried round-bottom flask under a nitrogen atmosphere, and dissolved in dry THF (3.0 ml). The solution was cooled in a dry ice/acetone bath to $-78^\circ$C, and \textit{n}-BuLi (2.5 M in hexanes, 0.053 ml, 0.131 mmol, 10 eq.) was added. The reaction was stirred for 15 minutes in the dry ice/acetone bath and 45 minutes at room temperature. Afterwards, it was cooled in an ice bath and acetyl chloride (0.056 ml, 0.788 mmol, 60 eq.) was added to the cooled reaction mixture and was stirred 5 minutes in the ice bath and finally 1 hour at room temperature. To the completed reaction, dichloromethane (5 mL) was added and the volatiles were removed under reduced pressure. The crude mixture obtained was purified with flash column chromatography (heptane/ether: 4/1) to give the product as a yellow solid in 75\% yield (11 mg, 9.9 $\mu$mol).

\textbf{$^1$H NMR (500 MHz, CD$_2$Cl$_2$):} $\delta$ 7.74 (d, J = 1.8 Hz, 2H), 7.45 (d, J = 8.3 Hz, 2H), 7.41–7.38 (m, 4H), 7.05 (s, 2H), 6.83 (dd, J = 8.3, 1.8 Hz, 2H), 6.77 (s, 2H), 3.97–3.92 (m, 4H), 1.95 (s, 6H), 1.94 (s, 18H), 1.83–1.74 (m, 4H), 1.69 (s, 18H), 1.07 (t, J = 7.4 Hz, 6H).

\textbf{$^{13}$C NMR (126 MHz, CD$_2$Cl$_2$):} $\delta$ 195.22, 152.25, 149.84, 149.66, 137.65, 137.06, 133.97, 132.41, 131.92, 128.91, 128.06, 127.95, 127.25, 125.07, 121.92, 121.51, 121.27, 118.35, 114.51, 114.38, 113.51, 104.31, 104.02, 45.35, 35.58, 35.02, 30.87, 30.27, 29.85, 23.17, 12.43.

\textbf{HR-MS (MALDI-TOF):} calcd.\ for C$_7$0H$_7$0N$_2$O$_5$S$_2$ [M]$^{\bullet +}$ is 1082.4726 m/z, found 1082.4736 m/z.

\subsubsection{NMR spectra}

\begin{figure}[t]
\centering
\includegraphics[width=0.9\linewidth]{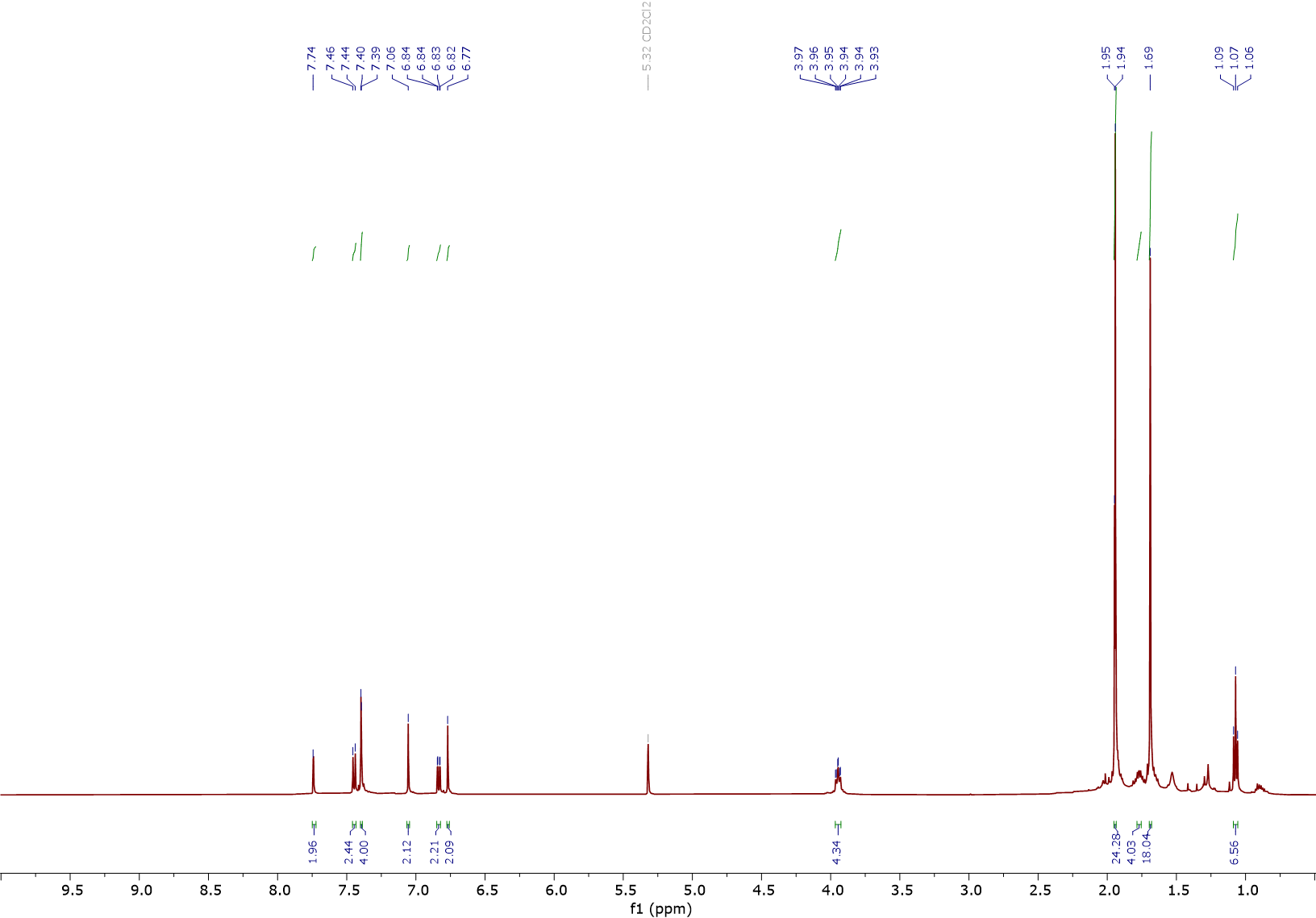}
\caption{$^1$H NMR spectrum of \textbf{thioacetyl[13]helicene} in CD$_2$Cl$_2$, 500 MHz.}
\label{fig:1H_NMR_heli}
\end{figure}

\begin{figure}[t]
\centering
\includegraphics[width=0.9\linewidth]{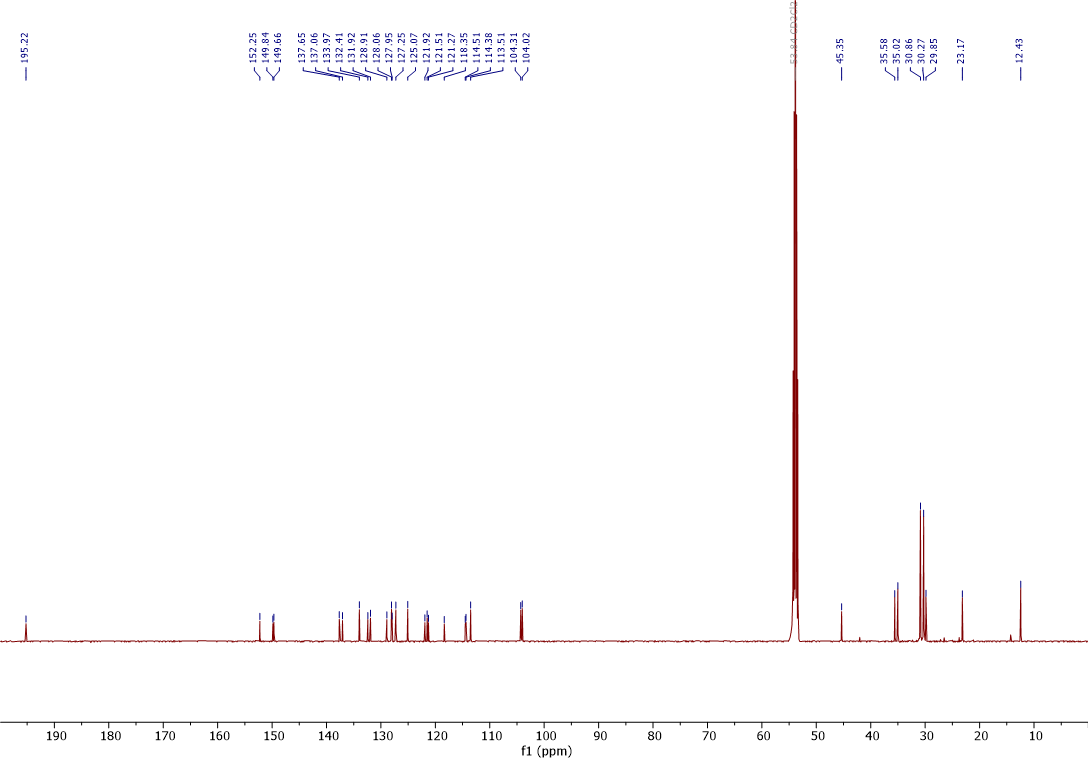}
\caption{$^{13}$C NMR spectrum of \textbf{thioacetyl[13]helicene} in CD$_2$Cl$_2$, 125 MHz.}
\label{fig:13C_NMR_heli}
\end{figure}

\begin{figure}[t]
\centering
\includegraphics[width=1\linewidth]{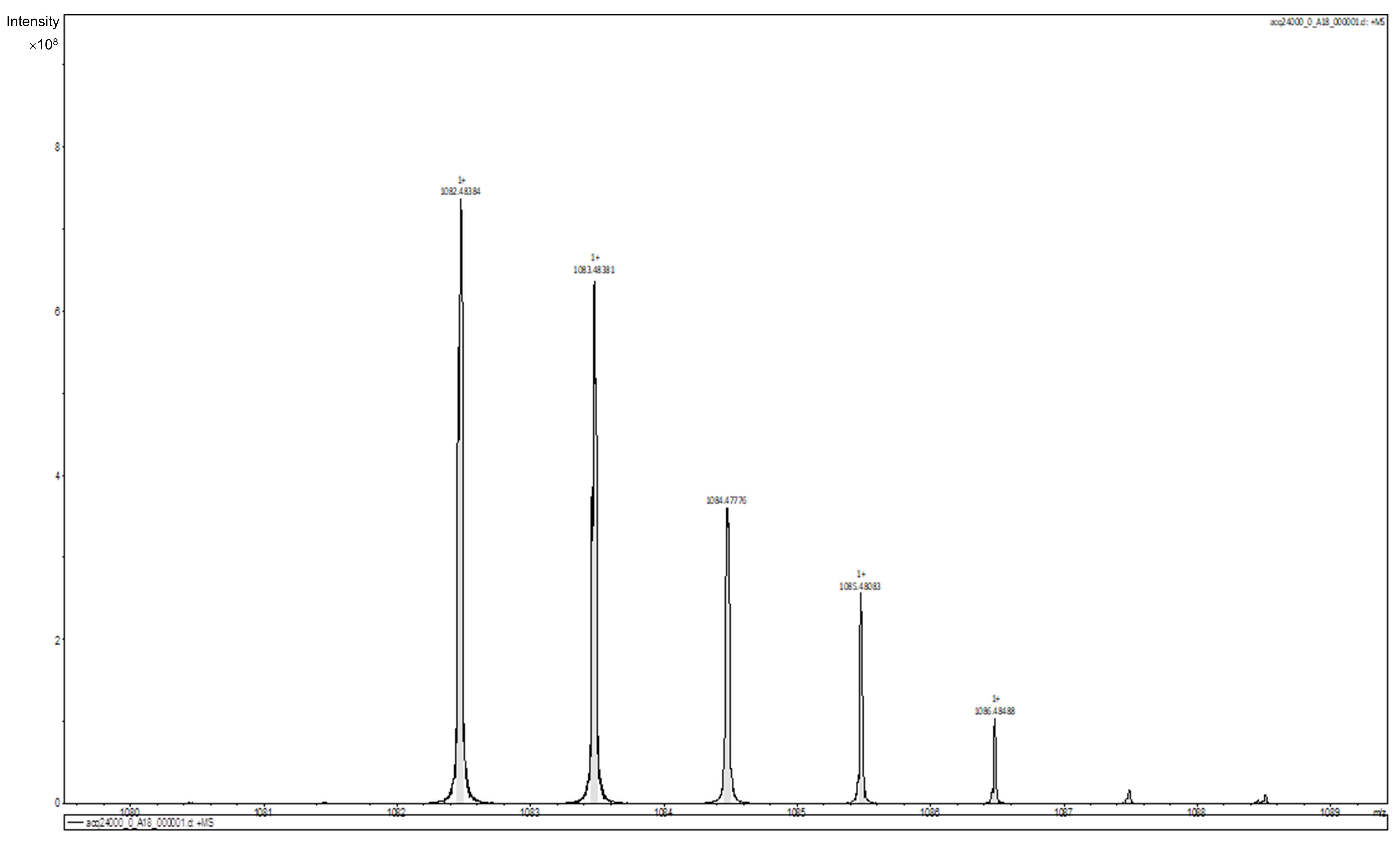}
\caption{ HRMS (MALDI‑TOF) spectrum of \textbf{thioacetyl[13]helicene }[M]\textsuperscript{\textsuperscript{•+} }.}
\label{fig:HRMS}
\end{figure}

\clearpage

\section{Electron transport measurements}

\subsection{Measurement setup}

The described measurements of electronic conductance as a function of interelectrode displacement in different molecular junctions are performed using a mechanical controllable break-junction (MCBJ) set-up at $\sim$4.2 K (Figure 1a, main text). The sample is fabricated by attaching a wire of Au (99.998$\%$, 0.1mm, Alfa Aesar) with a notch at its center to a flexible substrate (1-mm-thick phosphor-bronze plate covered by 100 $\mu$m insulating Kapton film). This structure is placed in a vacuum chamber and cooled by liquid helium. A three-point bending mechanism is used to bend the substrate and break the wire at the notch under cryogenic vacuum conditions. This procedure forms an adjustable gap between two freshly formed ultraclean atomically sharp tips. A piezoelectric element (PI P-882 PICMA) is used to fine-tune the bending of the substrate and control the distance between the electrodes with sub-Å resolution. The piezoelectric element is driven by a 24-bit NI-PXI4461 data acquisition (DAQ) card, connected to a piezo driver (Piezomechanik SVR 150/1).

An ensemble of junctions with diverse structures is studied by repeatedly compressing the electrodes together to form a contact of $\sim$50–70 $G_{0}$ ($G_{0}\cong1/12.9$ $k\Omega^{-1}$ is the conductance quantum) and pulling the electrodes apart until full rapture, using the piezo element, at a rate of 10–20 Hz, while simultaneously measuring the conductance. To measure conductance, the junction is biased with a d.c. voltage, provided by the NI-PXI4461 DAQ card. The presented measurements are performed at a bias voltage of 200 mV (2.0 V voltage is applied by the DAQ card, via a 1/10 voltage divider to improve the signal to noise ratio). The resulting current from the junction is amplified by a current amplifier (Femto amplifier DLPCA 200) and recorded by the DAQ card at a sampling rate of 100–200 kHz. To extract the conductance, the obtained current values are divided by the applied voltage value. Before the introduction of the molecules, we analyzed the typical conductance of the metal atomic junctions by recording conductance traces as a function of the relative electrode displacement.

\begin{figure}[t]
\centering
\includegraphics[width=1\linewidth]{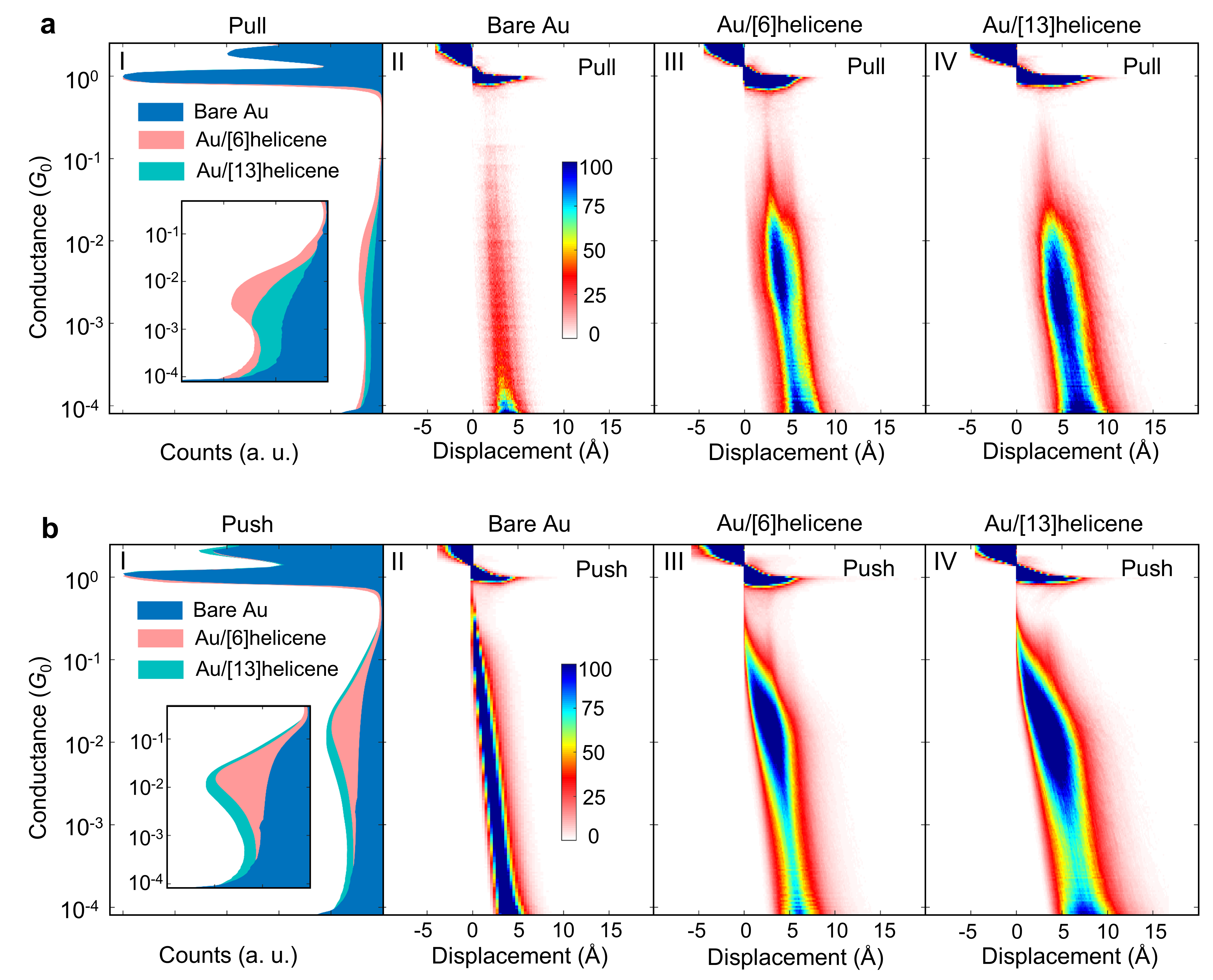}
\caption{Conductance histogram and conductance-displacement density plot of conductance traces taken during the pull and push processes. (\textbf{a}) Conductance histogram of bare Au (blue), Au/[6]helicene (red), and Au/[13]helicene (Cyan) (I) constructed from pull traces. Conductance-displacement density plots for bare Au (II), Au-[6]helicene-Au (III), and Au-[13]helicene-Au junctions constructed from pull traces. (\textbf{b}) Conductance histogram of bare Au (blue), Au/[6]helicene (red), and Au/[13]helicene (cyan) (I) constructed from push traces. Conductance-displacement density plots for bare Au (II), Au-[6]helicene-Au (III), and Au-[13]helicene-Au junctions constructed from push traces. Each histogram is based on 20,000 conductance-displacement traces, taken during the pull and push processes at an applied voltage of 200~mV.}
\label{FigureS1}
\end{figure}

The interelectrode displacement is determined using the exponential dependence of tunneling currents on the electrode separation, as detailed in Refs.~\citenum{Tamar-2013, Sudipto-2022} and their supplementary materials. The racemic mixture of [6]helicene (2,2’-dithiol-[6]helicene) molecules was synthesized following the procedure outlined in Ref.~\citenum{Anil-2024}, while a racemic mixture of [13]helicene (SAc[13]SAc) was prepared as described in section 1. These molecules are introduced in situ from a local heated molecular source into the cold atomic-scale Au junction during repeated junction pull-push cycles, where the two electrodes are pulled apart or squeezed toward each other, respectively.

\clearpage

\subsection{Characterization of Au/helicene molecular junctions}

To comprehensively characterize the key conductance features, we constructed one dimensional conductance histograms and conductance-displacement density plots from a dataset comprising 20,000 traces without any data selection, as illustrated in Figure \ref{FigureS1}. In Figure \ref{FigureS1}a (I), depicted in blue, peaks at $1G_{0}$ correspond to the most probable conductance of bare Au atomic junction \cite{Yanson-1998, Ohnishi-1998}, while the tail at low conductance signifies tunneling transport measured immediately after the Au contact rupture. The conductance-displacement density plots in Figure \ref{FigureS1}a (II) illustrate the conductance evolution during the elongation of the bare Au junctions. Following the introduction of the target molecules, Figure \ref{FigureS1}a showcases the conductance histograms of Au/[6]helicene (red) and Au/[13]helicene (cyan) junctions, respectively. While the peak at $1G_{0}$ remains evident, an additional peak at lower conductance emerges, indicating the formation of molecular junctions. Figures \ref{FigureS1}a (III) and (IV) show the conductance displacement density plots of Au/[6]helicene and Au/[13]helicene junctions, revealing new spots appearing below the typical conductance of the Au junction, indicative of molecular junction formation.

Among the 20,000 traces collected during the elongation of Au/[6]helicene and\\ Au/[13]helicene junctions, 70$\%$ and 62$\%$ of the traces, respectively, exhibit molecular features. The remaining traces depict only Au features without any apparent features in the tunneling regime. Notably, the most probable conductance of the Au/[13]helicene molecular junctions is somewhat lower than that of the Au/[6]helicene junctions, with values around $2 \times 10^{-3}$ $G_{0}$ and $6 \times 10^{-3}$ $ G_{0}$, respectively, as illustrated in Figure \ref{FigureS1}a (I) and its inset. Figure \ref{FigureS1}b presents conductance histograms and conductance-displacement density plots constructed from 20,000 push traces. In Figure \ref{FigureS1}b (I), the conductance histogram of Au (blue), Au/[6]helicene (red), and Au/[13]helicene (cyan) junctions are displayed. The peak at $1G_{0}$ remains apparent, with additional peaks observed at lower conductance for Au/[6]helicene and Au/[13]helicene, indicating the formation of molecular junctions. Figures \ref{FigureS1}b (II), (III), and (IV) depict the conductance displacement density plots of Au, Au/[6]helicene, and Au/[13]helicene junctions, respectively, highlighting new spots appearing below the typical conductance of the Au junction in Figure \ref{FigureS1}b (III) and (IV), which signify molecular junction formation. Among the 20,000 traces collected during the compression of Au/[6]helicene and Au/[13]helicene junctions, analysis reveals that 82$\%$ and 70$\%$ of the traces, respectively, exhibit molecular features. The remaining traces depict only Au features without any discernible features in the tunneling regime. Notably, the most probable conductance of the Au/[13]helicene and Au/[6]helicene molecular junctions during the compression process is the same, with values around $1 \times 10^{-2}$ $G_{0}$, as seen in Figure \ref{FigureS1}b (I). We suggest that the order of magnitude higher conductance for the [13]helicene junction in the compression process compared to the elongation process originates from the scarcity of the very extended configurations (less conductive) during the compression process, while during the elongation process more extended (less conductive) configurations are more common. This difference is less apparent for the shorter [6]helicene junctions as each round in the helix contributes to the extension capability and the latter contains one round less.

\subsection{U-shape conductance traces of Au/helicene molecular junctions}

\begin{figure}[t]
\centering
\includegraphics[width=0.95\linewidth]{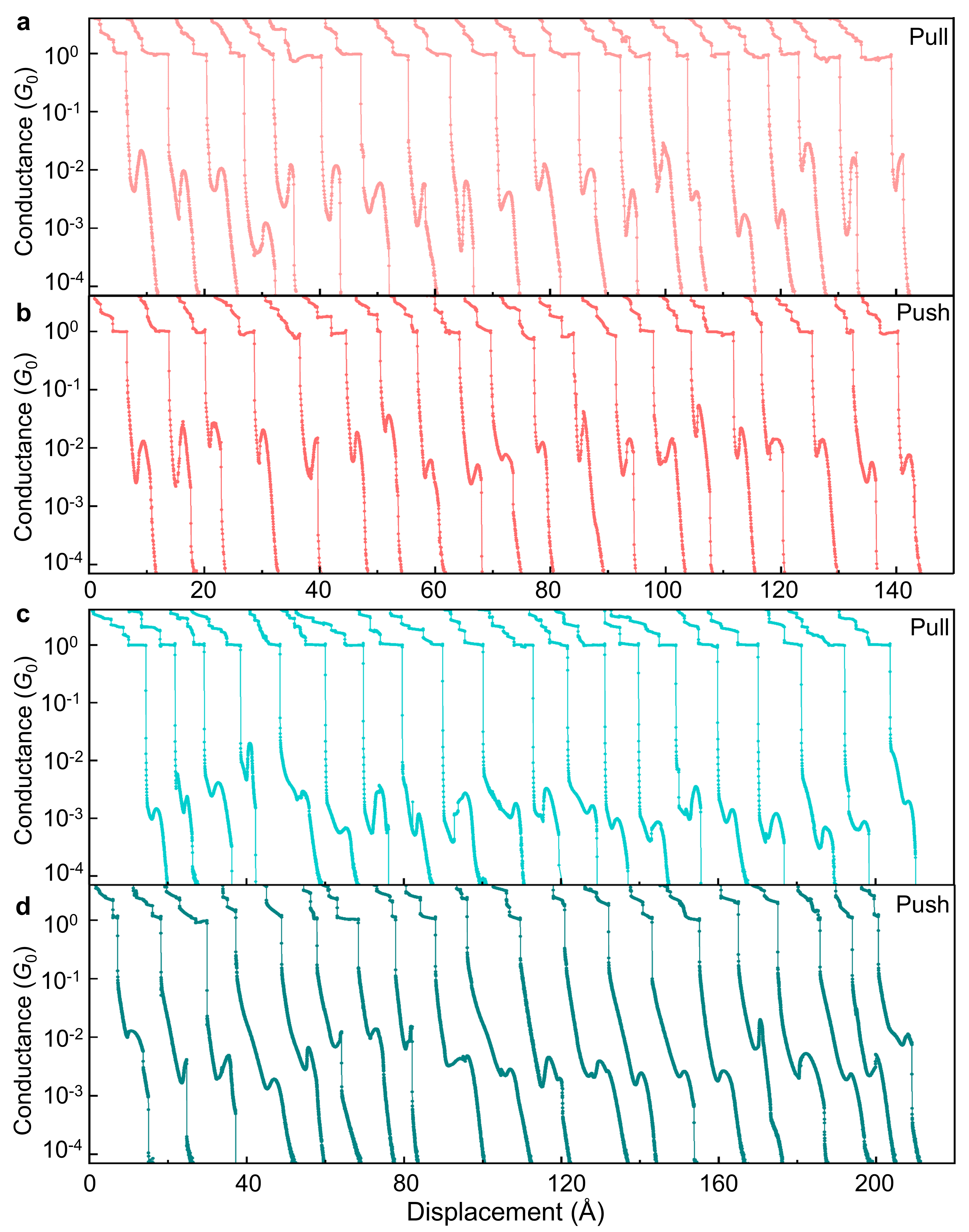}
\caption{Examples of conductance traces as a function of interelectrode displacement (shifted for clarity) for molecular junctions exhibiting a U-shape feature. (\textbf{a}) and (\textbf{b}) show conductance traces during the elongation and compression processes for Au/[6]helicene molecular junctions, respectively. (\textbf{c}) and (\textbf{d}) show similar data for Au/[13]helicene molecular junctions.}
\label{FigureS2}
\end{figure}

We have observed various manifestations of U-shape features in the collected molecular conductance versus interelectrode separation traces (denoted as molecular traces). These features occur at slightly different conductance values and exhibit somewhat different characteristics. To illustrate this, figure \ref{FigureS2} (a) and (b) display twenty examples of conductance traces with U-shape features during the elongation and compression processes for Au/[6]helicene molecular junctions, respectively. Figures \ref{FigureS2} (c) and (d) present similar data for Au/[13]helicene molecular junctions.

\subsection{Extraction of U-shape traces and conductance displacement density plot}

\begin{figure}[t]
\centering
\includegraphics[width=1\linewidth]{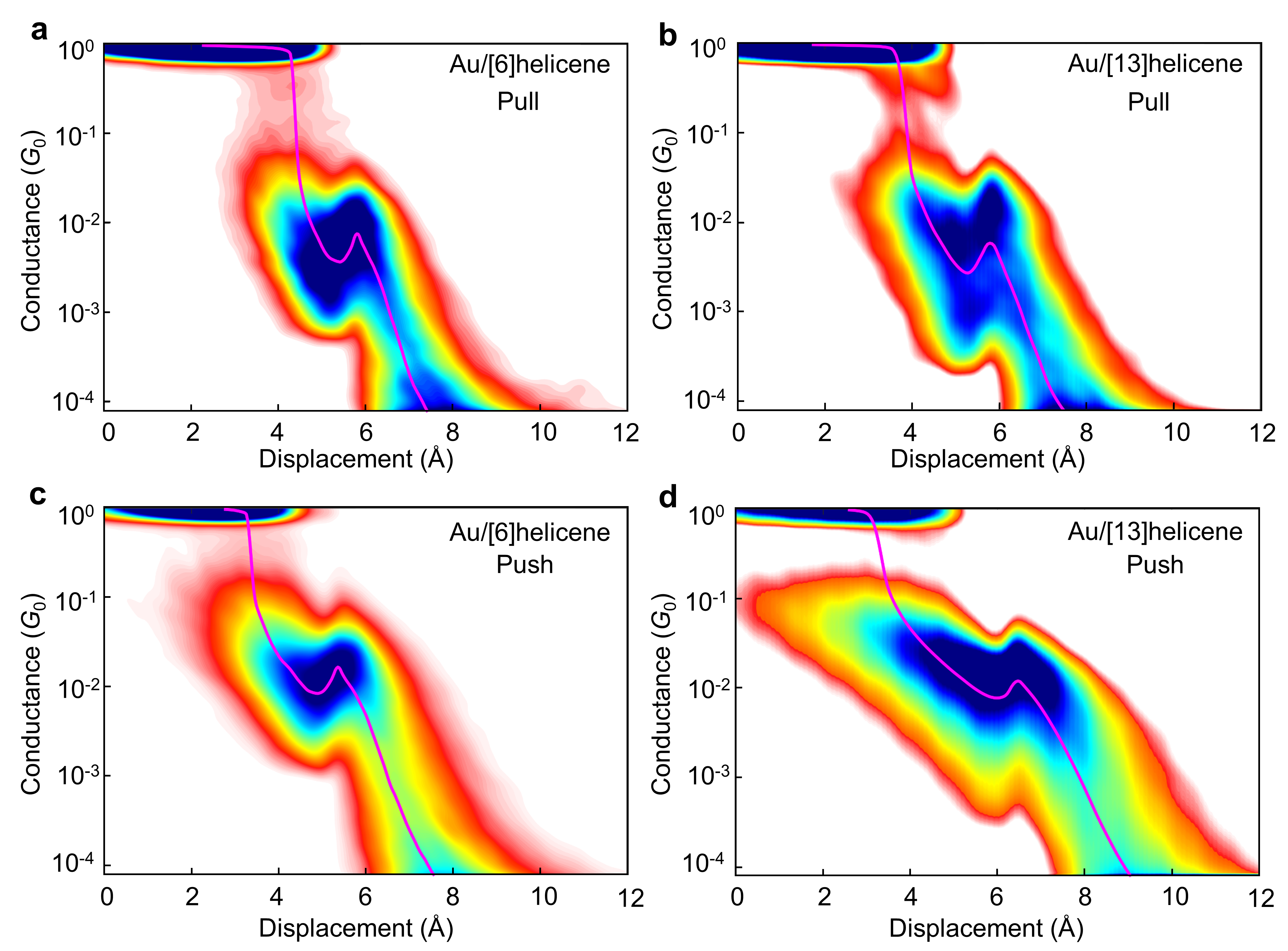}
\caption{Conductance-displacement density plot of conductance traces exhibiting a U-shape feature during the pull and push processes. (\textbf{a}) Conductance-displacement density plot for Au/[6]helicene molecular junction during pulling. (\textbf{b}) Conductance-displacement density plot for Au/[13]helicene during pulling. (\textbf{c}) Conductance-displacement density plot for Au/[6]helicene during the push process. (\textbf{d}) Conductance-displacement density plot for Au/[13]helicene during the push process. The magenta line in all plots represents the median trace.}
\label{FigureS3}
\end{figure}

To extract the conductance traces that have a U-shape feature out of a set of 20,000 raw traces, we first identify the U-shape features in the following unsupervised way. We begin by smoothing the traces using a 5-point moving average, followed by finding a local minimum and maximum by differentiating the conductance with respect to displacement. Traces with a local minimum that is followed by a local maximum at a higher interelectrode displacement (with a minimal difference of 4$\%$ in their conductance) are considered as traces with a U-shape feature. The corresponding raw traces (not smoothen) that contains a U-shape are then considered. The conductance-displacement density plots presented in Figure 2 of the main text are based on these traces. To better recognize the U-shape feature in the density plots of Figure 2, the individual traces that construct the plot are aligned at the local minimum of their U-shape feature. The sharpness at the minimum of the U-shape feature in the median traces presented in Figure 2 is a result of this alignment. Figure \ref{FigureS3} shows equivalent conductance-displacement density plots to the ones presented in Figure 2. However, in Figure \ref{FigureS3} the conductance traces are aligned at the position of the local maximum of the U-shape feature. In this case, a sharp feature at the local minimum of the median traces is not observed, but the maximum appears sharper.

To characterize the U-shape feature in the extracted conductance traces, we identify its starting and ending points in the tunneling regime, below the conductance typical to an atomic scale metallic contact. To identify the starting point, we examine the trace section that starts at the rupture of the atomic junction at 1 $G_{0}$ and ends at the local minimum. We then analyze its logarithmic slope (the derivative of logarithmic conductance with respect to length). This allows us to identify the starting point of the U-shape feature as a deviation from this slope by 20$\%$. Similarly, we determine the U-shape ending point by analyzing the segment of the conductance trace from the local maximum down to the measurement conductance noise floor at $8 \times 10^{-5}$ $G_{0}$ conductance, and examining the logarithmic slope of this segment. We identify the ending point of the U-shape feature as a deviation from this slope by 15$\%$. We then estimate the length of the U-shape feature and the percentage change in upturn conductance. These values are relevant for understanding the behavior of U-shape features in the molecular junctions. Figure 3 in the main text provides a detailed visual representation of these quantities. The one-dimensional conductance histograms in the figure illustrate the distribution of the U-shape feature lengths and the percentage changes in upturn conductance across the different traces. This analysis helps to quantify the variations and provides insights into the molecular conductance properties.

\clearpage

\section{Electronic structure and quantum transport calculations}

\subsection{Computational settings for electronic structure calculations}

All electronic structure calculations, reported in this publication, employed density functional theory (DFT), as implemented in the quantum chemistry package TURBOMOLE\cite{TURBOMOLE2023}. We chose PBE\cite{Perdew1996} as exchange-correlation functional and def-SVP \cite{Schafer1992} as basis set for all atoms. To take van der Waals interactions into account between stacking parts (for instance within the loops of the helicene molecules or at the helicene-electrode interfaces), a dispersion correction was included\cite{disp3}. The criterion for energy convergence in self-consistent field (SCF) iterations was set to $10^{-8}$~a.u.

\subsection{Modeling of junction geometries}

We model the central part of a single-molecule junction through an extended central cluster (ECC), which includes the molecule and part of the metal electrodes on left and right sides\cite{pauly2008cluster}, see figure~4 of the main text. Initial geometries are built by placing an isolated [6]helicene or [13]helicene molecule in its optimized geometry between two Au electrodes. For each of the two helicenes, we prepared two junction types with different electrode structures. In top-top (TT) junctions, the electrodes are atomically sharp pyramids with the transport direction oriented along the crystallographic (111) axis. In hollow-hollow (HH) junctions, tip atoms are removed, yielding blunt tips. Examples of the junction geometries at selected electrode separations are shown at the top of figure~4 in the main text.

Starting from on energetically optimized initial configuration, we stretch or compress the ECC by increasing or reducing the distance between the electrode clusters, respectively. Keeping the positions of the Au atoms in the two outmost layers on each side fixed, i.e.\ those Au layers that are most distant from the molecule, we separate or narrow these fixed layers in steps of 0.1~\AA. In every displacement step, we optimize the energy of the ECC geometry under the constraint of the fixed layers. This procedure is repeated, until the molecular junction ruptures on the long displacement end or until the DFT calculations do not converge anymore due to an unphysical compression on the short displacement end.
Molecular and ECC junction geometries are optimized until the norm of the Cartesian gradients falls below $10^{-5}$~a.u.

\subsection{Computational settings for quantum transport calculations}

For each junction geometry we obtain the electronic transmission using Landauer-B\"uttiker scattering theory, expressed through nonequilibrium Green’s function (NEGF) techniques\cite{pauly2008cluster}. Since experiments are conducted at low temperatures of $T\approx 4.2$~K, we determine the electrical conductance directly from the transmission through the relation
$$G=G_0\tau(E_\text{F})$$
with the conductance quantum $G_0=2e^2/h$. In figure~4 we decompose the total transmission at the Fermi energy into its eigenchannels,\cite{Cuevas:Book2017,pauly2008cluster} $\tau(E)=\sum_i \tau_i(E)$. In order to obtain converged transmission values, we use $32 \times 32$ transverse $k$ points in the quantum transport calculations\cite{pauly2008cluster}.

\begin{figure}[t]
\centering
\includegraphics[width=1\linewidth]{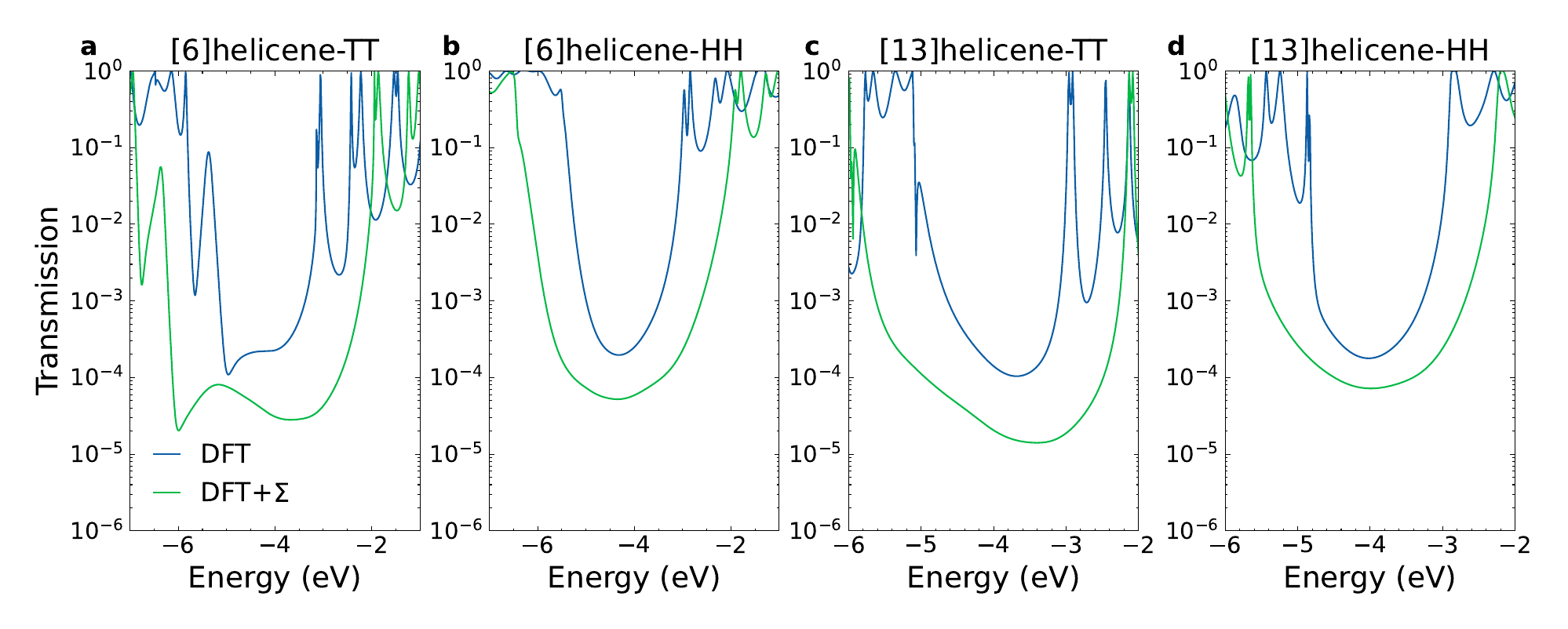}
\caption{Transmission as a function of energy for DFT and DFT+$\Sigma$ calculations. Single-molecule junctions are (a) [6]helicene in the TT geometry at an electrode separation of $d=3$~\AA, (b) [6]helicene in the HH geometry at an electrode separation of $d=4$~\AA, (c) [13]helicene in the TT geometry at an electrode separation of $d=3$~\AA, and (d) [13]helicene in the HH geometry at an electrode separation of $d=4$~\AA.}
\label{SI_dfts}
\end{figure}

A typical approach to improve transmission calculations based on DFT is the DFT+$\Sigma$ correction scheme\cite{dfts_neaton,dfts_zotti}. In order to avoid the computational costs of the DFT+$\Sigma$ method at every point of the stretching process, we mimic the effect of the correction of the molecule-electrode level alignment by adjusting the value of the Fermi energy $E_\text{F}$ in the original DFT transport calculations. For this purpose, we perform a DFT+$\Sigma$ calculation on one geometry for each type of junction studied. The comparison of DFT and DFT+$\Sigma$ transmission curves is shown in figure~\ref{SI_dfts}. As expected the gap between the resonances originating from HOMO and LUMO states is increased for DFT+$\Sigma$, and the transmission inside the HOMO-LUMO gap drops.

We adjust the Fermi energy $E_\text{F}$ in the DFT-based quantum transport calculations such that the ratio between the distances of the resonance peaks stemming from HOMO and LUMO frontier orbitals to the Fermi energy is the same as in the DFT+$\Sigma$-corrected results:\cite{Hsu-2022}
$$\frac{E_\mathrm{F}-E_\mathrm{HOMO}}{E_\mathrm{LUMO}-E_\mathrm{F}}=\frac{E_\mathrm{F,corr}-E_\mathrm{HOMO,corr}}{E_\mathrm{LUMO,corr}-E_\mathrm{F,corr}}.$$
The Fermi energy of the Au electrodes for junctions with applied correction is known to be\cite{pauly2008cluster} $E_\mathrm{F,corr}=-5$~eV.

This procedure yields Fermi energies of around $-4.7$~eV for [6]helicene junctions and $-4.5$~eV for [13]helicene junctions. These values are applied in figure~4 of the manuscript. The larger deviation from $-5$~eV in the [13]helicene junctions compared to the [6]helicene junctions agrees with the increased amount of carbon atoms in the ECC.

\subsection{Crossing of transmission eigenchannels}
In figure~4 of the manuscript we observe that the transmissions of eigenchannel 1 and 2 get very close to each other at selected displacements for the [6]helicene-TT and [13]helicene-HH junctions. These displacements are located near pronounced local minima of the total transmission. Since eigenchannels are ordered by the size of their transmission, we need to inspect the transmission eigenchannel wavefunctions in order to determine, if the eigenchannels cross at these points or if they just get very close without intersecting.

\begin{figure}[t]
\centering
\includegraphics[width=1\linewidth]{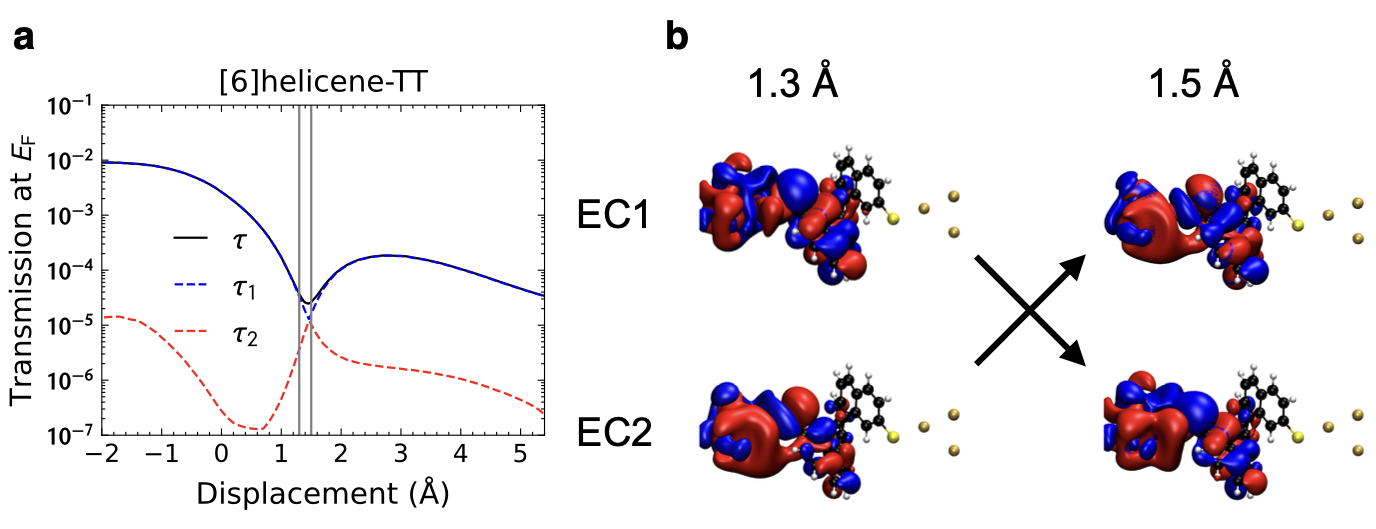}
\caption{(a) Total transmission, $\tau(E_\text{F})$, and those of the eigenchannels 1 and 2 with largest transmission, $\tau_{1}(E_\text{F})$ and $\tau_{2}(E_\text{F})$, as a function electrode displacement $d$ in the vicinity of the transmission minimum at $d=1.4$~\AA\ for the [6]helicene-TT junction. (b) Left-incoming wavefunctions of eigenchannels 1 (EC1) and 2 (EC2) at the Fermi energy for electrode displacements of $d=1.3$~\AA\ and 1.5~\AA\ demonstrate the exchange of characteristics and hence the crossing of channels. The gray vertical lines in panel (a) indicate the displacement points, where the eigenchannel wavefunctions are plotted. } 
\label{fig:ec_cross}

\end{figure}

To demonstrate that the channels indeed cross, we show the left-incoming wavefunctions of the first and second eigenchannel for the [6]helicene-TT junction in figure~\ref{fig:ec_cross}. Evaluated at the Fermi energy using the procedure explained in Ref.~\citenum{Burkle-2012}, we find that the character of the transmission eigenchannel wavefunctions is indeed exchanged after the displacement of $1.4$~\AA.

\subsection{Four-level model}

In figure~5 of the main text, we are presenting a four-level model that reproduces the essential features of the DFT-based two-dimensional energy-distance-dependent transmission map, shown in figure~4a. We will discuss technical details of this model in the following.

We consider a molecule coupled to two electrodes. The electronic transmission can be expressed by the Landauer formula as\cite{Cuevas:Book2017,pauly2008cluster}
$$\tau(E,d) = \mathrm{Tr}\left[\mathbf{\Gamma}_\text{L}(E,d) \mathbf{G}_\text{CC}^\text{r}(E,d) \mathbf{\Gamma}_\text{R}(E,d) \mathbf{G}_\text{CC}^\text{a}(E,d)\right].$$
Here $\mathbf{G}_\text{CC}^\text{r}$ denotes the retarded Green's function of the molecule, which is related to the advanced Green's function $\mathbf{G}_\text{CC}^\text{a}=(\mathbf{G}_\text{CC}^\text{r})^\dagger$ by complex conjugation and transposition, and $\mathbf{\Gamma}_X=-2\mathrm{Im}(\mathbf{\Sigma}^\text{r}_{X})$ is the linewidth broadening matrix of lead $X=\text{L},\text{R}$, which is derived from the retarded self-energy $\mathbf{\Sigma}^\text{r}_X$.

To simplify the expressions, we make use of the wide-band approximation, yielding energy-independent self-energies and linewidth broadening matrices. We further assume that both electrodes only interact with the terminal sulfur atoms, which establish the covalent bonds to the metal. For a linewidth broadening matrix of the form $\mathbf{\Gamma}_X=\gamma\mathbf{1}_x$, with a scalar constant $\gamma$ and the identity matrix $\mathbf{1}_x$ on the space of the orbitals of the left (l) or right (r) sulfur atom, $x=\text{l,r}$, matrix multiplication yields
$$\tau(E,d)/\gamma^2 = \text{Tr}_\text{l}[\lvert\mathbf{G}_\text{lr}^\text{r}(E,d)\rvert^2].$$
In the expression, $\mathbf{G}_\text{lr}^\text{r}(E,d)$ is the retarded Green's function describing propagation from left to right sulfur orbitals, and the subscript l at $\text{Tr}_\text{l}$ shows that the trace is taken only over the orbitals (or basis functions) at the left sulfur atom. The energy and distance dependence of the transmission now results basically from the Green's function of the molecule.

Simplifying even further by assuming that there is only a single relevant orbital on each sulfur, the Green's function becomes a scalar. If we then neglect the embedding self-energy, resulting from the coupling of the molecule to the electrode, the zeroth order retarded propagator is given as
$$ G_\text{lr}^{(0),\text{r}}(E,d)=\sum_m\frac{c_{\text{l}m}(d)c_{\text{r}m}^*(d)}{E-\epsilon_m(d) + i \eta}=\sum_m\frac{\rho_{\text{lr},m}(d)}{E-\epsilon_m(d) + i \eta}. $$
with $\epsilon_m(d)$ being the orbital energy, $c_{xm}(d)$ with $x=\text{l},\text{r}$ the expansion coefficient of the $m$-th molecular orbital at the left or right sulfur atom, and $\eta$ is an infinitesimal broadening parameter, preventing the Green's function from diverging at the orbital energies. To simplify the notation, we have introduced the residues $\rho_{\text{lr},m}(d)=c_{\text{l}m}(d)c_{\text{r}m}^*(d)$.

Motivated by the inverse energy dependence of the contribution of each molecular orbital to the overall value of the Green's function, we consider only the four frontier molecular orbitals HOMO-1, HOMO, LUMO and LUMO+1. The spectral representation of the Green's function is therefore given by
$$ G_\text{lr}^{(0),\text{r}}(E,d)=\sum_{m=\mathrm{HOMO-1}}^\mathrm{{LUMO+1}}\frac{\rho_{\text{lr},m}(d)}{E-\epsilon_m(d) + i \eta}, $$
and the electronic transmission of the four-level model is finally
$$ \tau(E,d)/\gamma^2 = \left\lvert G_\text{lr}^{(0),\text{r}}(E,d)\right\rvert^2 = \left\lvert\sum_{m=\mathrm{HOMO-1}}^{\mathrm{LUMO+1}}\frac{\rho_{\text{lr},m}(d)}{E-\epsilon_m(d) + i \eta}\right\rvert^2. $$
In the literature\cite{Li-2012, Schosser-2022, Nozaki-2017, Yoshizawa-2012, Stefani-2018} the orbital energies $\epsilon_m(d)$ typically get approximated by distance-dependent functions, and the residues are set to constant values of $\pm 1$.

Making use of the spectral decomposition of the Green's function, the standard orbital rule for electron transport through molecules\cite{Yoshizawa-2012, Li-2012, Koga-2012} can be deducted, as follows: Restricting the sum to just HOMO and LUMO as well as dropping the distance dependence, the zeroth-order retarded Green's function reads
$$ G_\text{lr}^{(0),\text{r}}(E)=\frac{\rho_\text{lr,HOMO}}{E-\epsilon_\text{HOMO}+i\eta}+\frac{\rho_\text{lr,LUMO}}{E-\epsilon_\text{LUMO}+i\eta}. $$
Since $E-\epsilon_\text{HOMO}$ and $E-\epsilon_\text{LUMO}$ have opposite signs at an energy $E$ inside the HOMO-LUMO gap, the two summands will have the same (opposite) signs, if the residues $\rho_\text{lr,HOMO}$ and $\rho_\text{lr,LUMO}$ have opposite (same) signs. The sign of $\rho_{\text{lr},m}=c_{\text{l}m}c_{\text{r}m}^*$ arises from the expansion coefficients of orbital $m$ at the particular sites $x=\text{l},\text{r}$, i.e.\ the shape of the wave function at the electrode-connecting sulfur atoms. Therefore this rule predicts a transmission valley or **destructive quantum interference** inside the HOMO-LUMO gap, if the signs of the residues $\rho_\text{lr,HOMO}$ and $\rho_\text{lr,LUMO}$ are the same.

The four-level model with residues restricted to $\rho_{\text{lr},m}=\pm 1$ confirms the orbital rule for electron transport\cite{Yoshizawa-2012,Nozaki-2017, Stefani-2018}, i.e., a destructive quantum interference occurs inside the HOMO-LUMO gap, if and only if residues of HOMO and LUMO possess the same sign \cite{Yoshizawa-2012}. A more complex behavior may however arise, if other values for the $\rho_{\text{lr},m}$ are allowed. As shown in figures 4 and 5 of the main text, transmission valleys can then occur, even if HOMO and LUMO levels do not have residues of equal signs. In this sense a **violation of the orbital rule** for electron transport occurs, if more than two levels are permitted with residues deviating from $\pm 1$. In particular, as we show in figure 5 of the main text, the four-level model actually predicts scenarios in which molecules feature destructive interferences inside the HOMO-LUMO gap even though $\rho_\text{lr,HOMO}$ and $\rho_\text{lr,LUMO}$ possess opposite signs.
With [6]helicene in the TT junction configuration, we have discovered such a molecular junction even in the DFT calculations, see figure~4. The orbital plots in figure~5b clearly show that the residues $\rho_\text{lr,HOMO}$ and $\rho_\text{lr,LUMO}$ have opposite signs, but still the transmission map displays a pronounced transmission valley.

Before addressing the numerics, let us provide some analytical insights. For this we bring the Green's function to the same denominator, resulting in
$$ G_\text{lr}^{(0),\text{r}}(E)= \frac{\sum_{m=\text{HOMO-1}}^\text{LUMO+1}\rho_{\text{lr},m}
\prod_{j\neq m}(E-\epsilon_j + i \eta)}{\prod_{n=\text{HOMO-1}}^\text{LUMO+1}(E-\epsilon_n + i\eta)}. $$
Now expanding the numerator, continuing to drop the distance dependence for brevity and calling the levels HOMO-1, HOMO, LUMO and LUMO+1 just 1 to 4 yields
$$ G_\text{lr}^{(0),\text{r}}(E){\prod_{n=1}^4(E-\epsilon_n + i\eta)}=f(E)+ i\eta g(E) - \eta ^2 h(E) - i \eta^3 (\rho_{\text{lr},1}+\rho_{\text{lr},2}+\rho_{\text{lr},3}+\rho_{\text{lr},4}). $$
Since $\eta$ is an infinitesimal parameter, solving $f(E,d)=0$ is a very accurate approximation for the course of destructive quantum interferences, as long as they do not get close to the transmission resonances resulting from the orbital energies. The function is given by
$$f(E,d) = \sum_{m=\text{HOMO-1}}^\text{LUMO+1}\rho_{\text{lr},m}(d)\prod_{j\neq m}(E-\epsilon_j(d)).$$
At $E_\text{HOMO}$ and $E_\text{LUMO}$ we obtain
$$ \text{sign}(f(E_\text{HOMO},d))= \text{sign}(\rho_\text{lr,HOMO}(d)), $$
$$ \text{sign}(f(E_\text{LUMO},d))= -\text{sign}(\rho_\text{lr,LUMO}(d)). $$

Relation for $f(E,d)$ shows that $f(E,d)$ is a polynomial of order three in energy at a given distance, which can therefore exhibit at most three zeros. If $\text{sign}(\rho_\text{lr,HOMO}(d))=\text{sign}(\rho_\text{lr,LUMO}(d))$, the sign change of $f(E,d)$ between $\epsilon_\text{HOMO}(d)$ and $\epsilon_\text{LUMO}(d)$, implies a zero point of the transmission in the HOMO-LUMO gap. If there is a single zero, this is the case covered by the standard orbital rule for electron transport, but there might also be three zeros (or a single zero of order three). If $\text{sign}(\rho_\text{lr,HOMO}(d))=-\text{sign}(\rho_\text{lr,LUMO}(d))$ instead, $f(E,d)$ has equal signs at $\epsilon_\text{HOMO}$ and $\epsilon_\text{LUMO}$. In this case, there might be no zero inside the HOMO-LUMO gap, as predicted by the standard two-level scenario of the orbital rule for electron transport, where at most a single zero can occur, but there could also be a double zero point (or a single zero of order two). This latter case implies that destructive quantum interferences outside of the "allowed area" of the standard two-level orbital rule for electron transport are possible and will typically appear in sets of two.

Generally speaking, the analysis shows that if we consider $n$ frontier orbitals in the spectral decomposition, up to $n-1$ zeros can occur inside the HOMO-LUMO gap at a fixed distance $d$. The course of the destructive quantum interferences as a function of energy and distance will strongly depend on the values of expansion coefficients and orbital energies. As compared to two levels with residues of $\pm 1$, a far more complex behavior of transmission valleys might occur at the cost of an increased amount of parameters.

The previous analysis gives rise to the following question: Why do we see only one transmission valley in the HOMO-LUMO gap for [6]helicene-TT in figure~4? The answer is that the solution of $f(E,d)=0$ only accurately approximates destructive quantum interferences, if they are not too close to an orbital energy. As it can be seen by the white dotted line in the transmission map of figure 5, which shows the course of the transmission valley for a sufficiently small $\eta$, approximating the solution $f(E,d)=0$, one of the destructive quantum interferences is located very close the LUMO. Because of this, the transmission valley is hidden by the corresponding resonance peak. In figure~4, the same effect is visible in the DFT calculations. There we track transmission dips of the first eigenchannel by searching for points, where the transmissions of first and second eigenchannels are similar.

The four-level model, shown in figure~5 of the main text, uses the actual orbital energies calculated from the DFT analysis. This improves the accuracy of the model. But a similar transmission map could also be generated for constant energy levels. We therefore conclude that for the helicene single-molecule junctions, the orbital energies do not matter so much, when the course of the transmission valleys is studied.
The part of the four-level model that is thus crucial for the course of the destructive interferences in helicene junctions are the residues or orbital expansion coefficients. They are functions of $d$, since the relevant orbitals are rotated in response to variations in displacement, as shown in figure~5 of the main text.

\begin{figure}[t]
\centering
\includegraphics[width=1\linewidth]{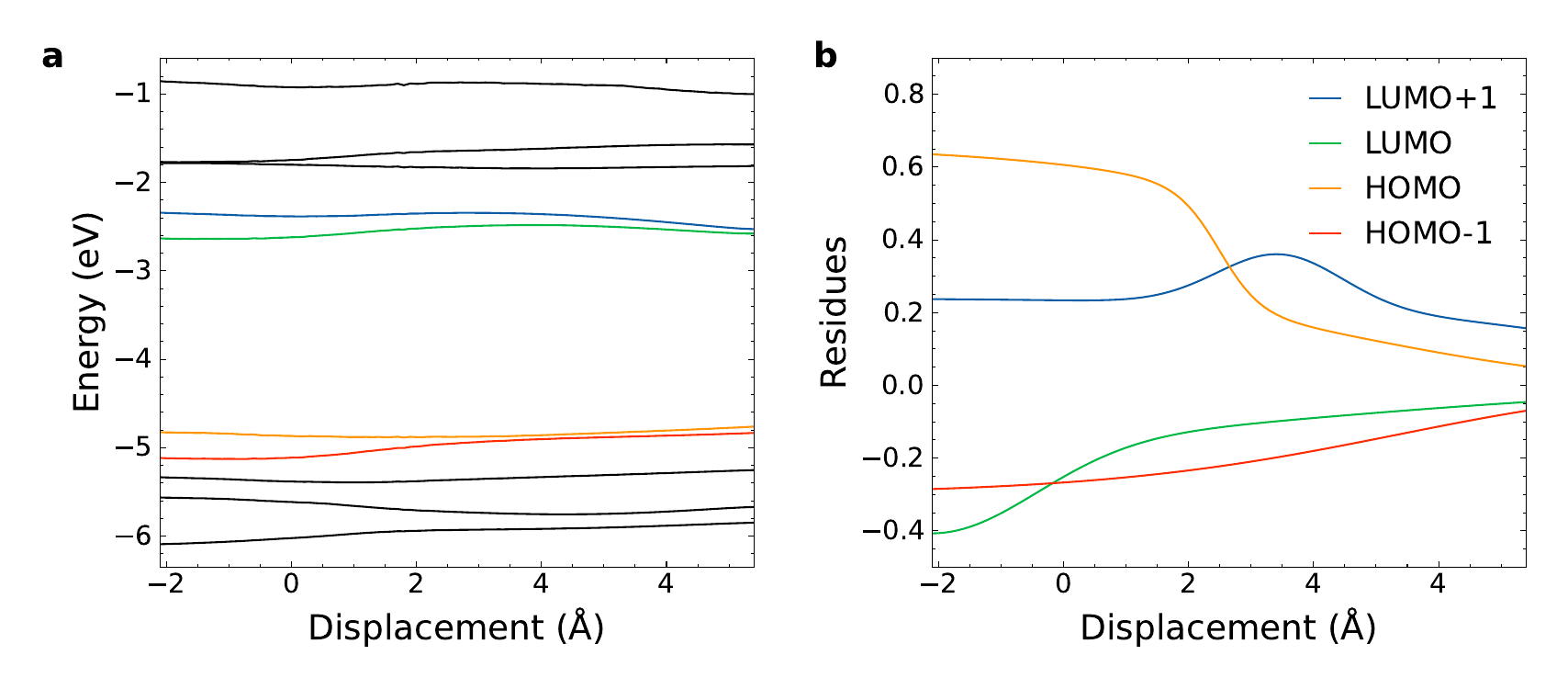}
\caption{(a) Evolution of the energies of molecular orbitals $\epsilon_m$ as a function of electrode displacement. The values are computed with DFT for geometries of the isolated [6]helicene molecule, obtained from the [6]helicene-TT junction by removing all Au atoms and saturating the terminal sulfur atoms by a single hydrogen atom at each end. The four molecular orbitals considered in the four-level model are plotted in color. (b) Residues $\rho_{\text{lr},m}$ of the molecular orbitals $m=\text{HOMO-1}, \text{HOMO}, \text{LUMO}, \text{LUMO+1}$ as a function of electrode displacement. The values are chosen such as to reproduce the two-dimensional DFT-based transmission map of the [6]helicene-TT junction in figure~4a with the four-level toy model, whose transmission is shown in figure~5c.}
\label{fig:S4}
\end{figure}

Figure~\ref{fig:S4}a shows the orbital energies of the isolated [6]helicene molecule, as calculated from the DFT simulation. The geometries of the isolated molecule are obtained from the [6]helicene-TT junction by removing the Au atoms of the ECC and saturating the sulfur anchors on each side with a single hydrogen atom. Since HOMO-1, HOMO and LUMO, LUMO+1 are closest to the HOMO-LUMO gap and occur as a pair, we just consider these levels, which are colored in the plot. As shown by figure~5c, the four levels are sufficient to explain the DFT-based transport results of figure~4a.

Figure~\ref{fig:S4}b displays the choice of the residues $\rho_{\text{lr},m}$ as a function of electrode separation for the four frontier orbitals, i.e.\ $m=\text{HOMO-1}, \text{HOMO}, \text{LUMO}, \text{LUMO+1}$. Comparing the values of the expansion coefficients with the two-dimensional transmission map of figure~5c in the main text, it becomes clear that the transmission valley gets induced by the decrease in absolute value of the residue of the LUMO, which drops below those of the LUMO+1 around a displacement of $d=0$~\AA. At this point the transmission valley arises. The two zeros of the transmission split up, and the suppressed transmission close to the LUMO is hardly visible due to its vicinity to this resonance. The lower zero shows up as the well-visible transmission valley, which then travels towards the HOMO resonance as a result of a decreasing residue of the HOMO.

The peculiar occurrence of two zeros of the transmission inside the HOMO-LUMO gap at a fixed electrode displacement can be explained by analyzing the Green's function of the four-level model. Actually, for explaining two zeros, only three states need to be considered. We discuss the situation exemplarily at the case relevant for figure~5. We assume that (i) the signs of $\rho_\text{lr,HOMO}$ and $\rho_\text{lr,LUMO}$ are different, i.e.\ $\text{sign}(\rho_\text{lr,HOMO})=-\text{sign}(\rho_\text{lr,LUMO})$. According to the orbital rule for electron transport for two levels no destructive quantum interference would then be expected inside the HOMO-LUMO gap. However if the following two further conditions apply, two destructive quantum interferences, showing up as two zeros of the energy-dependent transmission function, will emerge in the HOMO-LUMO gap: (ii) The absolute value of the residue of the LUMO is significantly smaller than the residue of the LUMO+1, (iii) the signs of these residues differ, i.e.\ $\text{sign}(\rho_\text{lr,LUMO})=-\text{sign}(\rho_\text{lr,LUMO+1})$. The term "significantly smaller" needs to be put into relation to the distance in energy space between LUMO and LUMO+1. For a smaller energetic difference between LUMO and LUMO+1 a smaller difference in residues is sufficient to induce the two destructive interferences.

Let us now discuss the position of zeros inside the HOMO-LUMO gap. The bigger the difference of $|\rho_\text{lr,LUMO}|$ and $|\rho_\text{lr,LUMO+1}|$, the closer will the lower zero be to the HOMO and the upper zero to the LUMO. Thus the separation of zeros will increase in energy. The splitting between the two zeros can be seen as a stability criterion for observing destructive quantum interferences. For a large splitting, the occurrence of two zeros is more robust to variations in orbital energies and residues than for a small splitting.
We note that the zero at lower energies can be assigned to the interference of the HOMO and LUMO+1, the one close to the LUMO to the interference of LUMO and LUMO+1.

Two destructive quantum interferences inside the HOMO-LUMO gap at a fixed electrode displacement can obviously also arise, if we exchange the role of HOMO and LUMO in the description above and if additionally the HOMO-1 takes the role of the LUMO+1.

If both the (HOMO-1,HOMO) pair as well as the (LUMO,LUMO+1) pair fulfill the conditions (i) to (iii) above, i.e. $\text{sign}(\rho_\text{lr,HOMO})=-\text{sign}(\rho_\text{lr,LUMO})$, $\text{sign}(\rho_\text{lr,HOMO-1})=-\text{sign}(\rho_\text{lr,HOMO})$, $\text{sign}(\rho_\text{lr,LUMO+1})=-\text{sign}(\rho_\text{lr,LUMO})$ and $|\rho_\text{lr,LUMO+1}|$ sufficiently larger than $|\rho_\text{lr,LUMO}|$ as well as $|\rho_\text{lr,HOMO-1}|$ sufficiently larger than $|\rho_\text{lr,HOMO}|$, then two zeros of the squared propagator will be found in the HOMO-LUMO gap, one close to the HOMO, the other one close to the LUMO. Due to the large splitting these two destructive quantum interferences are in principle expected to be particularly robust. However, care should be taken, since they may not be perceivable due to broad transmission resonances. From this reasoning, we see that variations of orbital energies and residues with electrode displacement, as visible in figure~\ref{fig:S4}, can be used to switch from zero to two destructive quantum interferences. Additionally, the splitting of the two zeros can be controlled, and one or even both transmission minima can effectively be hidden by approaching HOMO and LUMO transmission resonances, see figure~5c of the main text.

Only one transmission eigenchannel exists in our four-level model due to the single relevant orbital assumed at the terminal sulfur atoms of the helicene molecules, see the scalar nature of the propagator. For this reason, in the four-level model the total transmission equals the transmission of the first eigenchannel, i.e.\ $\tau(E,d)=\tau_1(E,d)$. In contrast, in the quantum transport results from DFT calculations in figure~4 several transmission eigenchannels can contribute. The zeros of an effective four-level model, corresponding to suppressions of $\tau_1(E,d)$, may then be detected in the DFT results by searching for energy and distance points, where transmissions of first and second eigenchannels are similar, i.e.\ $\tau_1(E,d)\approx\tau_2(E,d)$.

\clearpage

\bibliography{bibliography}